\documentclass[%
 reprint,
 superscriptaddress,
 amsmath,amssymb,
 aps,
]{revtex4-1}

\usepackage[english]{babel}
\usepackage[utf8]{inputenc}
\usepackage{color}
\usepackage[colorlinks=true, allcolors=blue]{hyperref}
\usepackage{dcolumn}
\usepackage{bm}
\usepackage[caption=false]{subfig}
\usepackage{graphicx}
\usepackage{amsmath}
\usepackage{braket}
\usepackage{bbold}
\usepackage[normalem]{ulem}
\usepackage{dsfont}

\DeclareMathOperator{\order}{\mathcal{O}}
\DeclareMathOperator{\Tr}{Tr}
\DeclareMathOperator{\Id}{\mathbb{1}}

\begin{document}

\title{Quantum East model: localization, non-thermal eigenstates and slow dynamics}

\author{Nicola Pancotti}
\affiliation{Max-Planck-Institut f\"ur Quantenoptik, D-85748 Garching, Germany}
\affiliation{Munich Center for Quantum Science and Technology (MCQST), Schellingstr. 4, D-80799 M\"unchen}
\affiliation{Department of Physics, Technische Universit\"at M\"unchen, 85748 Garching, Germany}
\author{Giacomo Giudice}
\affiliation{Max-Planck-Institut f\"ur Quantenoptik, D-85748 Garching, Germany}
\affiliation{Munich Center for Quantum Science and Technology (MCQST), Schellingstr. 4, D-80799 M\"unchen}
\author{J. Ignacio Cirac}
\affiliation{Max-Planck-Institut f\"ur Quantenoptik, D-85748 Garching, Germany}
\affiliation{Munich Center for Quantum Science and Technology (MCQST), Schellingstr. 4, D-80799 M\"unchen}
\author{Juan P. Garrahan}
\affiliation{School of Physics and Astronomy, University of Nottingham, Nottingham, NG7 2RD, United Kingdom}
\affiliation{Centre for the Mathematics and Theoretical Physics of Quantum Non-equilibrium Systems, University of Nottingham, Nottingham NG7 2RD, UK}
\author{ Mari Carmen Ba\~nuls}
\affiliation{Max-Planck-Institut f\"ur Quantenoptik, D-85748 Garching, Germany}
\affiliation{Munich Center for Quantum Science and Technology (MCQST), Schellingstr. 4, D-80799 M\"unchen}

\begin{abstract}
We study in detail the properties of the quantum East model, an interacting quantum spin chain inspired by simple kinetically-constrained models of classical glasses.
Through a combination of analytics, exact diagonalization and tensor-network methods we show the existence of a transition, from a fast to a slow thermalization regime, which manifests itself throughout the spectrum.
On the slow side, by exploiting the localization of the ground state and the form of the Hamiltonian, we explicitly construct a large (exponential in size) number of non-thermal states which become exact finite-energy-density eigenstates in the large-size limit, 
as expected for a true phase transition. 
A ``super-spin'' generalization allows us to find a further large class of area-law states proved to display very slow relaxation. These states retain memory of their initial conditions for extremely long times. 
Our numerical analysis reveals that the localization properties are not limited to the ground state and that many eigenstates have large overlap with product states and can be approximated well by matrix product states at arbitrary energy densities.  
The mechanism that induces localization to the ground state, and hence the non-thermal behavior of the system, can be extended to a wide range of models including a number of simple spin chains.
We discuss implications of our results for slow thermalization and non-ergodicity more generally in disorder-free systems with constraints and we give numerical evidence that these results may be extended to two dimensional systems.
\end{abstract}

\maketitle

\section{Introduction}

The dynamics and thermalization of interacting quantum systems are extremely challenging problems that attract considerable attention due to their fundamental and practical relevance to many areas of physical sciences, including condensed matter, quantum information, statistical mechanics and beyond (see \cite{Eisert2014,DAlessio2016} for reviews).
Despite many advances in the last couple of decades \cite{Eisert2014,DAlessio2016}, even for models as simple as one-dimensional spin chains with local interactions it has not been possible to reach a fully satisfactory and general understanding of these problems, neither through the use of analytical tools nor through numerical methods.
The main obstacles~\cite{Schuch2008, Osborne_2012, Calabrese_2005, Vidmar2017, Ashton_2006} relate to the growth of quantum correlations, the spreading of information, and the highly entangled nature of the excited eigenstates that dominate the dynamical evolution.

A central focus of research in the last decade has been the search for many-body systems with dynamics that falls outside the general paradigm of thermalizing quantum systems.
Undoubtedly, the most prominent example of this new class of interacting systems is that of those undergoing many-body localization (MBL)~\cite{Basko2006, Oganesyan2007} (see \cite{Nandkishore2015,Altman2015,Abanin2019,Gopalakrishnan2019} for reviews).
Inspired by the formidable analytical, numerical, and experimental advances in MBL, see e.g.~\cite{Imbrie2016,Berkelbach2010,Canovi2011, Bardarson2012, Serbyn2013, Huse2014, Andraschko2014, Laumann2014, Serbyn2014, Yao2014, DErrico2014, Serbyn2014, Schreiber2015, Choi2016, Znidaric2008, Nanduri2014, Kjall2014, Vosk2014}, more recently there has been a shift of emphasis towards the study of systems which also display non-thermal behavior but in the absence of quenched disorder.

The range of these comprises the search for MBL-like physics in translationally invariant or disorder-free models~\cite{Carleo2012,DeRoeck2014,Grover2014,Schiulaz2015,Papic2015,
Barbiero2015,Yao2016,Smith2017,Mondaini2017,Yarloo2018,Schulz2019,Nieuwenburg2019}, slow thermalization in systems with dynamical constraints which are either explicit \cite{Horssen2015,Hickey2016,Shiraishi2017,Lan2018,Feldmeier2019} or emergent (as in ``fractons'' \cite{Chamon2005,Haah2011,Castelnovo2012,Yoshida2013,Prem2017,Nandkishore2018, Khemani2019Local,khemani2019localization,Rakovszky2020,Sala2020,Pretko2020}), the existence of localised (almost) conserved operators (or ``strong zero modes'') in clean systems with boundaries \cite{Fendley2012,Fendley2016,Kemp2017,Else2017,Vasiloiu2019},
and the appearance of ``quantum scars'' \cite{Turner2018,Turner2018b,James2019,Ho2019,Ok2019,Schecter2019, Khemani2019Signatures, Hudomal2019} and other non-thermal excited eigenstates in otherwise thermalizing systems \cite{Znidaric2013, Moudgalya2018Exact, Moudgalya2018Entanglement}.

Here we address several of these questions by studying in detail the properties of the {\em quantum East model}, introduced in Ref.~\cite{Horssen2015} as a candidate disorder-free system displaying breakdown of ergodicity at long times, 
and further studied with and without disorder in the context of MBL in Ref.~\cite{Crowley2017,Crowley2019}. 
This model is inspired by the classical stochastic East model \cite{Jackle1991}, a prototypical kinetically constrained model (KCM) of classical glasses (for reviews on classical KCMs and their application to the glass transition problem see \cite{Ritort2003,Garrahan2018}).
The numerical simulations of Ref.~\cite{Horssen2015} suggested a possible transition in the quantum East model from a thermalizing phase where relaxation is fast, to a phase of slow relaxation where dynamics retains memory of initial conditions for long times indicating the possible absence of ergodicity.
However, as it is often the case with numerics for the small systems accessible to exact diagonalization, it is difficult to make convincing extrapolations from the results of \cite{Horssen2015} for the asymptotic behavior for large system sizes in the quantum East model.

We describe a novel mechanism that gives rise to non-thermal behavior in a broad class of interacting quantum systems. This mechanism is distinct from that of other constrained models such as the PXP~\cite{Turner2018} or quantum dimers~\cite{Lan2018,Feldmeier2019}.
For technical convenience we consider the case of open boundary conditions.
We employ a combination of analytical arguments,
exact diagonalization and tensor network methods to show that the model
displays a {\em fast-to-slow phase transition} throughout its spectrum,
by which we mean a change from a dynamical phase where thermalization is fast to a phase where dynamics is slow and even non-ergodic depending on initial conditions.
The transition occurs when changing the parameter that controls the balance between kinetic and potential energies in the Hamiltonian across a ``Rokhsar--Kivelson'' (RK) point \cite{Rokhsar1988,Castelnovo2005}.

In particular, we demonstrate that the slow dynamical phase is characterized by the following:
(i)~the ground state is exponentially localized and can be efficiently approximated for large system sizes;
(ii)~there is an exponentially large (in system size) number of non-thermal eigenstates at finite energy density that are non-thermal, which we show how to construct analytically for large system sizes by exploiting the localization of the ground state and the kinetically constrained nature of the Hamiltonian. This construction is very simple, i.e. a tensor product of two eigenstates of the same Hamiltonian supported on smaller sizes; (iii)~of these, at least a number which is linear in size has area-law entanglement, while for the rest their bipartite entanglement is spatially heterogeneous;
(iv)~these non-thermal eigenstates have large overlap with product states and can be approximated well by matrix product states (MPS) at arbitrary energy densities and large system sizes;
(v)~it is possible to generalize the construction to an even larger number of area-law states, i.e. tensor products of localized blocks or {\em super-spins}, that are guaranteed to display 
 very long memory of their initial conditions, exponential in the size of the block (super-spin). Accordingly, the time required to entangle a block is also exponential in its size.
(vi)~Extensive numerical analyses performed with exact diagonalization and tensor networks, reveal atypical dynamical properties of the model beyond the analytical constructions. The statistical study of several quantities of interest confirms the singular change throughout the spectrum, and suggest that our results may be further extended. In particular, we find that the localization properties of the ground state \--- which are cornerstones of our analytic results \--- are present for several excited states as well.

We prove, furthermore, that the mechanism which induces localization of the ground state can be extended to a large class of models and, numerically, we show that it may be present also in two dimensions.
As most of the non-thermal properties of the quantum East model arise from the localization of the ground state and they do not rely on the particular form of the Hamiltonian, we can deduce that all these generalizations will exhibit a similar atypical dynamical behavior.

The remarkable range of non-thermal features that we uncover here underlines the potential richness of non-equilibrium behavior of quantum KCMs with appropriately tailored constraints.

The paper is organized as follows.
In Sec.~\ref{sec:model} we introduce the quantum East model and describe its basic properties.
Section~\ref{sec:ground_state} considers the localization transition in the ground state of the model.
In Sec.~\ref{sec:exact_eigs} we propose an analytic ansatz in the localized/slow phase which allows us to construct an approximation to the ground state of a larger system starting from the exact ground state of a smaller system, and which becomes exact in the large size limit.
As a generalization of this procedure, we show how to analytically construct an exponential number in system size of approximate non-thermal eigenstates with finite energy-density using as ingredients eigenstates of smaller systems.
These become exact eigenstates in the large size limit.
Some of these states fulfill area-law of entanglement and hence can be efficiently approximated by MPS for large system sizes.

In Sec.~\ref{sec:frankenstein} we construct a large class of area-law states with small energy variance in terms of localized ``super-spins''.
While these are not strict eigenstates, unitary dynamics starting from these states is very slow, and we provide bounds to the decay of time-correlation functions and the growth of entropy with time.
In Sec.~\ref{sec:eigenstates}, we analyze in detail the statistical properties of the spectrum of small systems accessible to exact diagonalization, showing that the fast/slow transition is manifested in a change of eigenstate characteristics --- including their entanglement, localization and closeness to product states --- throughout the spectrum.
In Sec.~\ref{sec:generalization} we summarize all our results and we discuss the implications to quantum constrained dynamics more broadly, as well as generalizations to higher dimensions. We further compare our findings with other constrained dynamical models and highlight the main differences. 
Finally, we enumerate in Sec.~\ref{sec:future_directions} some possible new research directions.

\begin{figure}
  \hspace*{-0.5cm}
  \includegraphics[scale=0.65]{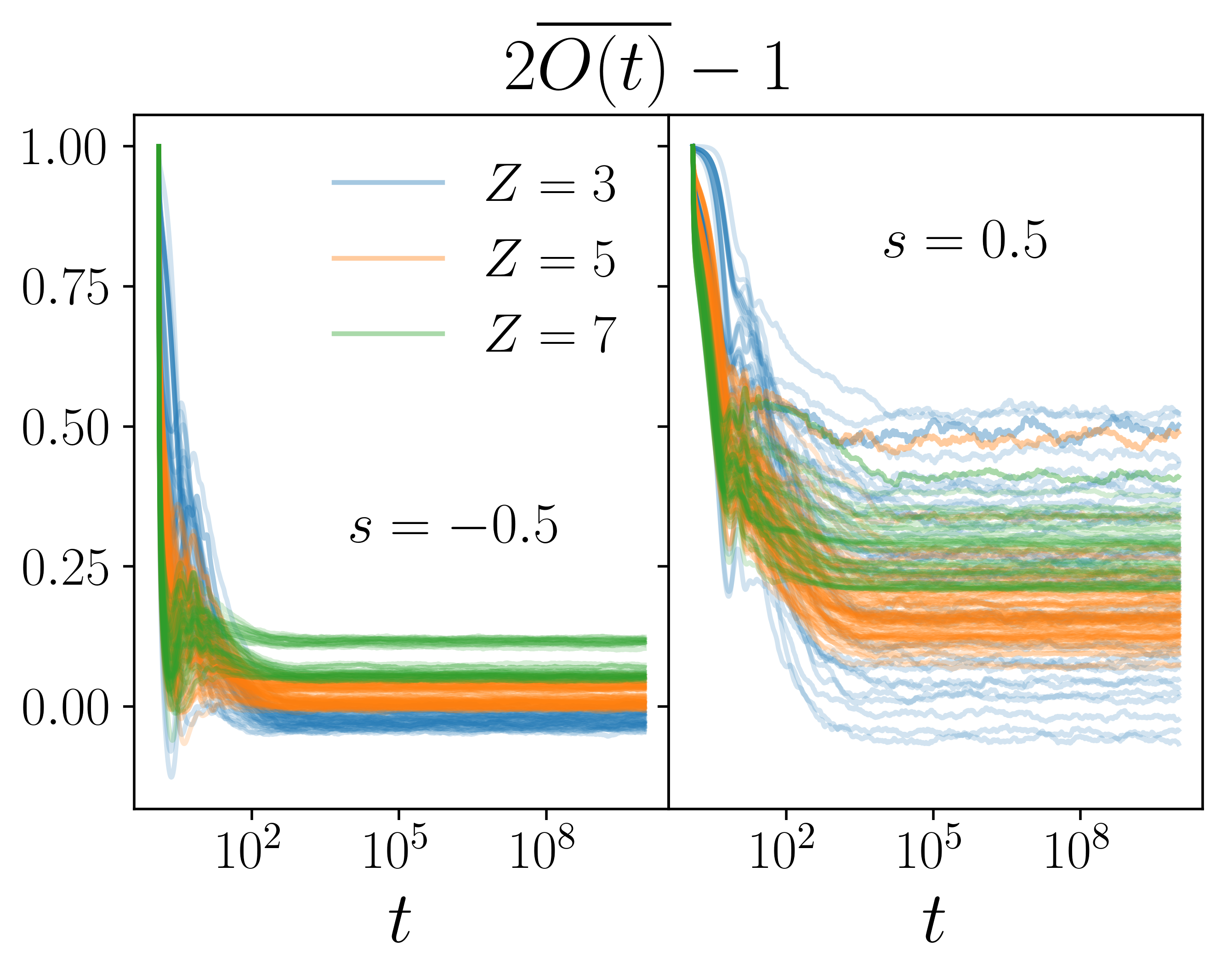}
  \caption{{\bf Fast vs.\ slow dynamics in the quantum East model.}
  Relaxation to equilibrium of (time-averaged) density autocorrelator \eqref{eq:autocorrelator}
  starting from all possible product initial states for $N=10$.
  For $s<0$ (left) equilibration is fast and memory of the initial conditions is rapidly lost.
  For $s>0$ (right) relaxation is slow and memory of initial conditions is preserved throughout the simulation window.
  }
  \label{fig:time_evolve_prod}
\end{figure}

\section{Quantum East model}\label{sec:model}

The quantum East model was originally introduced in \cite{Horssen2015} in order to consider slow quantum dynamics akin to (quasi-)MBL in the absence of disorder with kinetic constraints as the mechanism behind slow evolution.
The model is defined in terms of spin-$1/2$ degrees of freedom on a one dimensional lattice with Hamiltonian \cite{Horssen2015},
\begin{equation}\label{eq:QEast}
    H = - \frac{1}{2} \sum_{i=0}^{N} n_i (e^{-s} \sigma^x_{i+1} - \Id).
\end{equation}
where the operator $n_i = (\Id - \sigma^z_i)/2$ is a projector onto the state $\ket{1}$ in the local $z$-basis, and $\sigma^{\alpha}_i$ is the Pauli-$\alpha$ operator at site $i$.
When $s=0$, the operator in Eq.~\eqref{eq:QEast} is (up to a sign) the same as the continuous-time Markov generator of the classical East model, a stochastic KCM much studied in the context of the classical glass transition \cite{Jackle1991,Sollich1999,Garrahan2002,Faggionato2012,Chleboun2013}.
For $s \neq 0$, it corresponds to the ``tilted'' generator studied in the context of the dynamical large deviations of the stochastic model, see e.g.~\cite{Garrahan2007,Garrahan2018,Banuls2019}.
When considered as a quantum Hamiltonian, $s=0$ is a so-called RK point \cite{Rokhsar1988,Castelnovo2005}, where the ground state corresponds to the equilibrium probability of the stochastic model.

When interpreted as a stochastic generator, the operator Eq.~\eqref{eq:QEast} corresponds to the ``infinite temperature'' classical East model.
Note that this terminology does not refer to the temperature of the quantum system, but to the characteristics of the equilibrium probability, i.e., the ground state of Eq.~\eqref{eq:QEast} at the stochastic point $s=0$.
At infinite temperature, the equilibrium probability is uniform for all configurations, while at finite temperature the equilibrium state is not the equal weight combination of all configurations, see e.g. Ref.~\cite{Banuls2019}.

The factor $n_i$ at the front of each term in the Hamiltonian \eqref{eq:QEast} is the {\em kinetic constraint}.
It represents an operator valued rate, which in the case of $H$ above makes the action of the local Hamiltonian at site $i+1$ non-trivial only when $n_i$ projects into the state $\ket{1}$.
In the KCM jargon, when this constraint is satisfied, the site $i$ is said to ``facilitate'' dynamics of its $i+1$ neighbour (i.e., the one to the East, thus the name of the model) \cite{Ritort2003,Garrahan2018}.
In contrast to Ref.~\cite{Horssen2015}, here we will study the properties of the Hamiltonian \eqref{eq:QEast} with open boundary conditions.
We do this for technical convenience, as we do not expect the physics we uncover below to be very different for the case with periodic boundaries.

The key numerical observation in Ref.~\cite{Horssen2015} was the change in the dynamical behavior when the parameter $s$ is changed from one side of the RK point, that is from $s<0$, to the other side, that is $s>0$.
In Fig.~\ref{fig:time_evolve_prod} we reproduce this observation for the case of open boundaries: we show the relaxation to equilibrium of the normalized two-time density autocorrelator $2\overline{O(t)} -1$, defined as the time average $\overline{O(t)} = \frac{1}{t}\int_0^t O(t')dt'$ of
\begin{equation}\label{eq:autocorrelator}
O(t) \equiv \frac{1}{Z}\sum_i \braket{ n_i (t) n_i (0) },
\end{equation}
where $n_i(t)$ is the occupation operator in the Heisenberg picture under unitary evolution generated by the Hamiltonian Eq.~\eqref{eq:QEast}, and $Z \equiv \sum_i \braket{ n_i (0) }$ is a normalization factor for the initial occupation.
The figure shows results for initial states which are product states in the occupation basis (i.e., local $z$-basis) at different initial fillings (note that magnetization is not conserved in this model).
Notice that, for finite systems, the energy is determined not only by the initial polarization $Z$, but also by the occupation of the last site.
This is the reason why we observe two different thermal values for the same polarization.
This effect vanishes in the thermodynamic limit.

We observe two fundamentally different behaviors of the autocorrelator depending on the sign of $s$.
For $s<0$ dynamics is {\em fast} and most of the information about the initial state is quickly erased, as expected from thermalization and compliance with ETH \cite{DAlessio2016}.
In contrast, for $s>0$ dynamics is {\em slow} and for a large class of initial product states, memory of the initial conditions is retained at arbitrarily long times.
This is indicative of a transition in the quantum dynamics of the system.

Motivated by these results, in the following we will analyze the structure of the eigenstates of the Hamiltonian in order to collect information about the dynamical properties of the model both for finite system sizes and in the thermodynamic limit.

\subsection{Symmetries of the quantum East model}

Since the Hamiltonian is identically zero on the empty string $\ket{0\dots0}$,
for open boundary conditions the Hilbert space splits in blocks that are not connected by the dynamics.
Each block is determined by the position of the first occupied site, i.e. the $k$-th block corresponds to the subspace spanned by all classical configurations that start with a string of $k-1$ zeroes followed by a $1$.

In the following, we will mostly focus on the dynamics of a single block, with $N$ (dynamical) sites to the right of the first occupied one.
The position of the latter naturally introduces an edge, and the effective Hamiltonian on the $N$ dynamical sites to its right reads
\begin{equation}\label{eq:QEast_block}
  H^N = - \frac{1}{2} (e^{-s} \sigma^x_{1} - \Id) -\frac{1}{2} \sum_{i=1}^{N-1} n_i (e^{-s} \sigma^x_{i+1} - \Id).
\end{equation}
Since  $[H^{N},\sigma^x_{N}]=0$, the Hamiltonian in Eq.~\eqref{eq:QEast_block} can be further divided in the sum of two commuting terms $H^N = H_+^{N-1} \otimes \Pi_N^+ + H_-^{N-1} \otimes \Pi_N^- $, where $\Pi^{\pm} = (\Id\pm\sigma^x)/2$ are single site projectors onto $\ket{\pm}=(\ket{0}\pm\ket{1})/\sqrt{2}$, the eigenstates of $\sigma^x$,  and
\begin{equation}\label{eq:H+/-}
  \begin{split}
    H_{\pm}^{N-1} = &- \frac{1}{2}  (e^{-s} \sigma^x_{1} - \Id) - \frac{1}{2}  \sum_{i=1}^{N-2} n_i (e^{-s} \sigma^x_{i+1} - \Id) \\
    &- \frac{1}{2} n_{N-1} (\pm e^{-s} - 1).
  \end{split}
\end{equation}
In the rest of the paper we will study and discuss the properties of the Hamiltonians in Eq.~\eqref{eq:QEast}, \eqref{eq:QEast_block} and \eqref{eq:H+/-}.

\subsection{The special case $s=0$}

At the RK point, $s=0$, the Hamiltonian \eqref{eq:QEast} has an additional symmetry.
It can be written as a sum of projectors $H=\sum_i n_i \otimes \Pi^-_{i+1}$ which, in addition to the empty string, annihilates also a string of $\ket{+}$ states.
Thus the Hilbert space splits further in blocks determined by the lengths $m$ and $n$ of, respectively, the leading empty string and the trailing string of $\ket{+}$.
Hence the eigenstates have the form $ \ket{ 0 }^{\otimes m} \ket{1} \ket{\psi^{N_B}} \ket{-}\ket{+}^{\otimes n}$, where $N_B$ is the length of the dynamical part of the $(m,n)$-block, and $\ket{ \psi^{N_B}}$ is an eigenstate of the corresponding effective Hamiltonian,
\begin{equation}\label{eq:red_ham}
H^{N_B}_{s=0}= \Pi^-_1 + \sum_{i=1}^{N_B-1} n_i \otimes \Pi_i^- + n_{N_B}.
\end{equation}

\section{Ground state localization phase transition} \label{sec:ground_state}

\begin{figure}
	\hspace*{-1.0cm}
	\subfloat{
		\begin{picture}(0,0)
		\put(37,64){\includegraphics[scale=0.3]{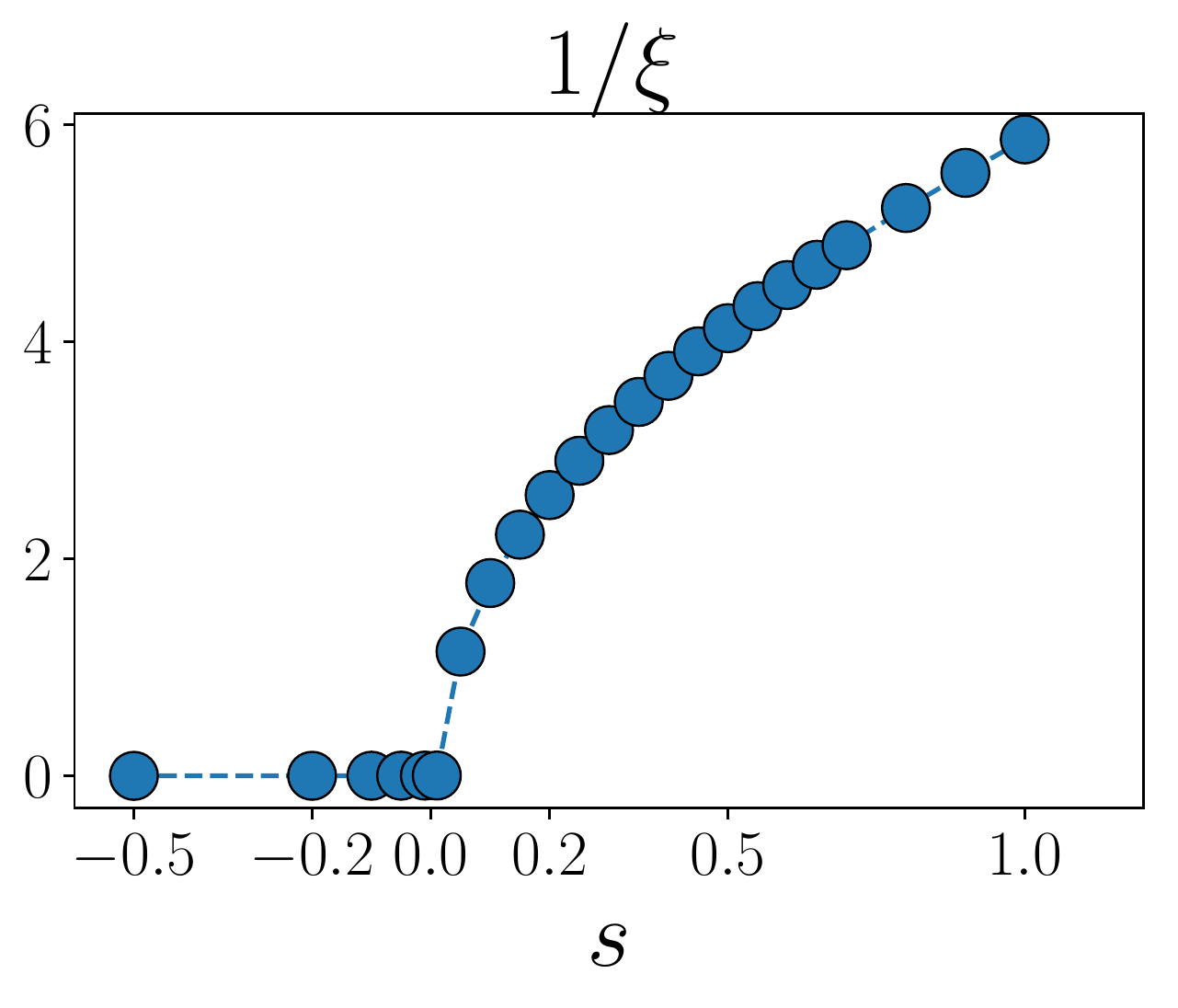}}
		\end{picture}
	}
	\includegraphics[width=0.5\textwidth]{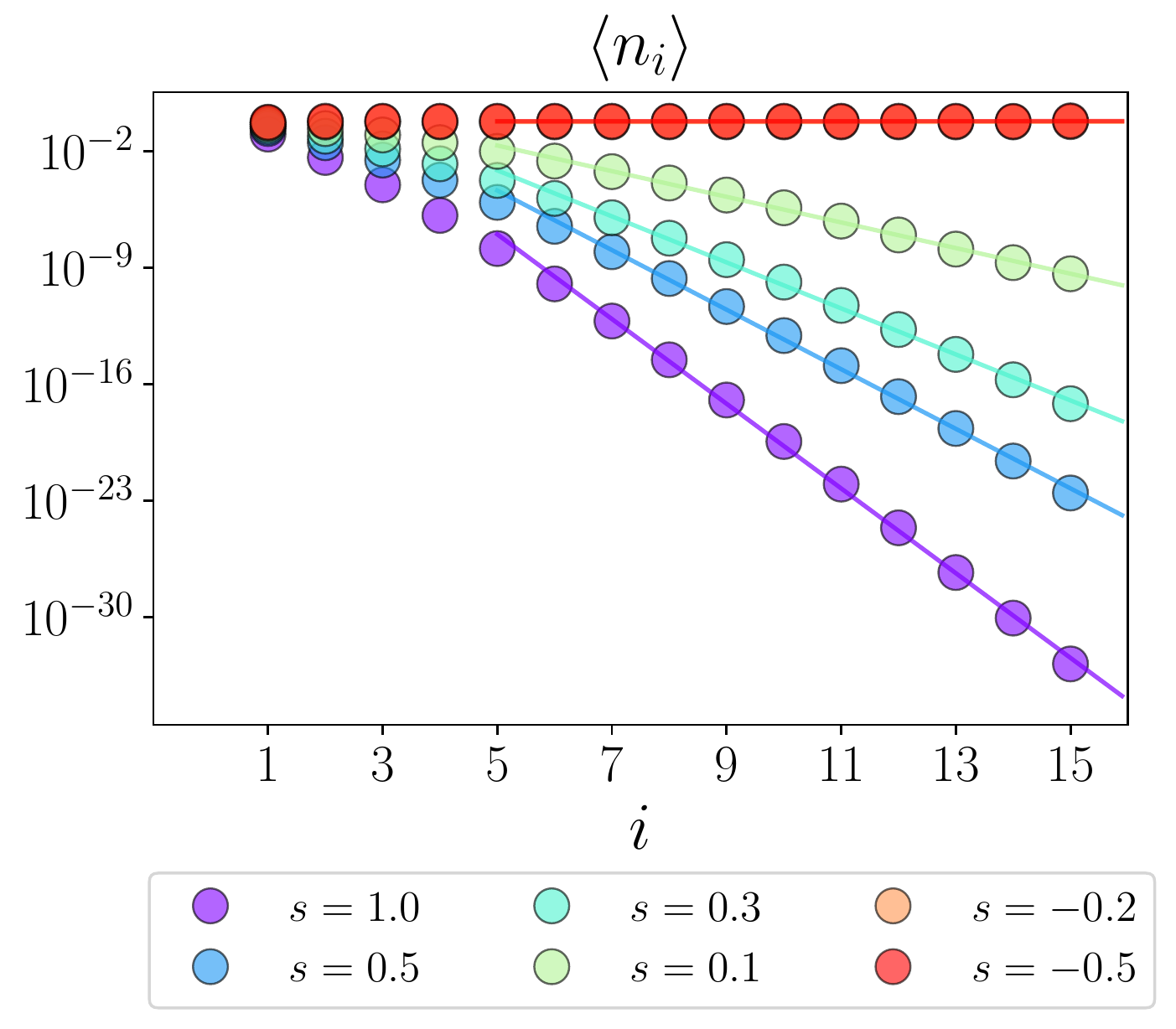}
	\caption{{\bf Localization of the ground state for $N=15$.} The main plot shows the single site occupation $\braket{n_k}$ as a function of the position in the chain $k$.
  For positive values of $s$ the probability of finding an occupied spin is exponentially suppressed as the distance from the left edge increases.
  {\bf Inset:} We fit the function~\eqref{eq:loc_length} and plot the inverse of the localization length $\xi$ as a function of the control parameter $s$.
	}
	\label{fig:loc_GS}
\end{figure}

\begin{figure}
  \hspace*{-0.5cm}
  \includegraphics[width=0.45\textwidth]{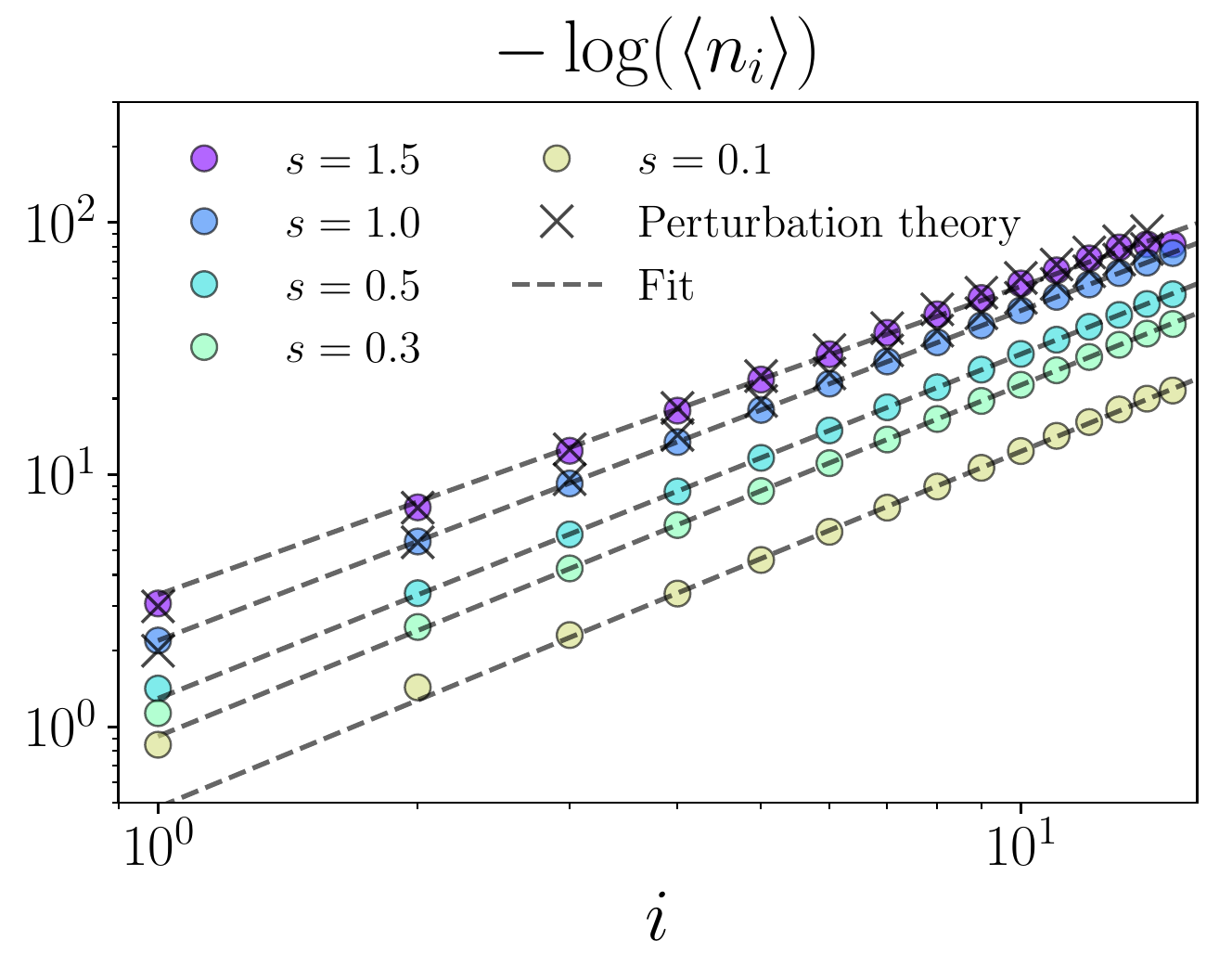}
  \caption{{\bf Super-exponential localization of the ground state.} We plot the behavior of $\braket{n_k}$ in double-log scale versus log scale.
  A fit of the form $\log \langle n_k \rangle \propto k^{\alpha}$ (dashed lines) yields $\alpha \sim 1.3$, indicating a super-exponential decay.
  The result from a perturbation theory calculation at large $s$ (crosses), without free parameters, shows good agreement with the numerical data.
  }
  \label{fig:super_exp_decay}
\end{figure}

\begin{figure}
  \includegraphics[width=0.35\textwidth]{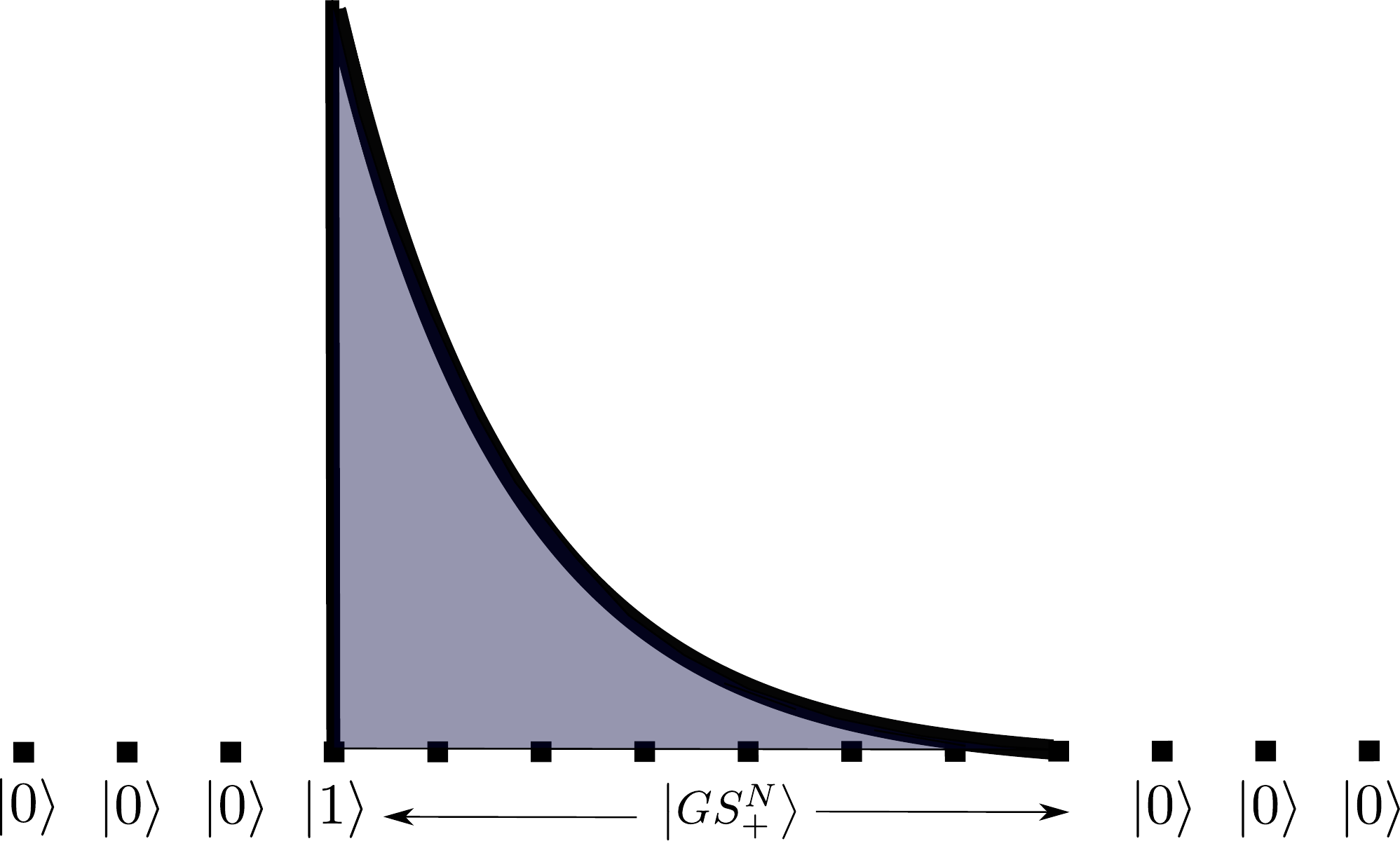}
  \caption{{\bf Sketch of a super-spin.} For positive values of $s$ the ground state is exponentially localized near the first occupied spin.
  The empty sites on the left of $\ket{1}$ are not dynamical.
  If we complete the state with $\ket{0}$s on the right, we obtain a good approximation of the ground state of the Hamiltonian~\eqref{eq:QEast} in the thermodynamic limit.
  }
  \label{fig:super_spin}
\end{figure}

We now show that the ground state of the quantum East model~\eqref{eq:H+/-} is localized when $s>0$.
Namely, in the ground state of a block of (dynamical) size $N$, the probability of finding an occupied site is exponentially localized in the neighborhood of a certain position and the state becomes a trivial product state further away, as depicted in Fig.~\ref{fig:super_spin}.
The localization length $\xi$ can be extracted already at small sizes, accessible by exact diagonalization, by analyzing the expectation value in the ground state of the local operator $n_k$ as a function of the position $k$.
This is shown in Fig.~\ref{fig:loc_GS} for the ground state of $H^N_+$, with $N=15$.
For $s<0$ we observe an almost homogeneous occupation, independent of the system size and the value of $s$.
For $s>0$, in contrast, the occupation decays fast with the distance to the edge, with faster decay as we increase $s$.
We find that the results can be fitted assuming an exponential decay,
\begin{equation}\label{eq:loc_length}
  \langle n_i \rangle \sim e^{-i/\xi},
\end{equation}
and the localization length $\xi$ from the fit captures the phase transition at $s=0$.
Indeed, we find that the value of $\xi$ diverges as $s=0$ is approached from the positive side, according to $\xi\propto s^{-\nu}$ with $\nu = 0.533 \pm 0.006$  (see inset of Fig.~\ref{fig:loc_GS}).
These results hold for the ground state of $H_+^N$ in Eq.~\ref{eq:H+/-}.
We observe the same qualitative behavior for the ground state of $H_-^N$.
Indeed, both Hamiltonians differ only in the last site, $H_+^N - H_-^N = - e^{-s} n_N$, with the difference decreasing fast for $s>0$.

The form in Eq.~\eqref{eq:loc_length} provides a good fit of the numerical data for the occupation, but a more detailed look at our numerical results suggests in fact a faster-than-exponential decay, as shown in Fig.~\ref{fig:super_exp_decay}.
Indeed, in appendix~\ref{appdx:perturbation_theory} we show that  for large $s$ perturbation theory provides an approximate decay of the form $\braket{n_i} \sim  (e^{-is}/i!)^2 \sim e^{-i\log(i)}$.
As can be seen in Fig.~\ref{fig:super_exp_decay}, this is in good agreement with the numerical data.

In Fig.~\ref{fig:super_spin} we provide a cartoon picture of the ground state which for $s>0$ is localized near the edge.
The spatial structure of the GS revealed by these studies can be understood in light of the adiabatic theorem.
Away from the phase transition, which happens at $s=0$~\footnote{For finite systems, the transition is actually shifted to a small $s_c>0$, which converges to $0$ faster than $1/N$ \cite{Banuls2019}.}, the system is gapped, and we can apply the adiabatic theorem to connect the ground state to the non-interacting one at $s\to\infty$.
The latter corresponds to the product state with only the first site occupied, $\ket{10\ldots 0+}$.
Within the gapped region, the evolution with the adiabatically changing Hamiltonian will dress the initial site with an exponential tail like the one shown in our numerical results and depicted in Fig.~\ref{fig:super_spin}.

This phenomenon is not exclusive to the quantum East model. As we discuss in Sec.~\ref{sec:generalization} and we demonstrate in appendix~\ref{appdx:generalization_localization}, there is a generic class of constrained Hamiltonians, including~\eqref{eq:QEast} as particular case, for which the ground state is exponentially localized.

\section{Eigenstates for large system sizes}\label{sec:exact_eigs}

Given the localization properties of the ground state discussed above, and the peculiar form of the Hamiltonian, in this section we provide an ansatz for the ground state and some excited states of finite energy density at arbitrarily large system sizes. These constructions rely on few simple assumptions and they are not limited to the quantum East model. Indeed, each Hamiltonian that belongs to the class defined in appendix~\ref{appdx:generalization_localization}, will show similar properties.

\subsection{The ground state for large system sizes}\label{subsec:exact_GS}

Consider the normalized state
\begin{equation}\label{eq:approx_gs}
  \ket{ \Psi_0 (L; N) } =  \ket{ {\rm GS} _{+} ^{N -1 } } \otimes \ket{ 0 } ^{\otimes (L-N+1)},
\end{equation}
where $\ket{ {\rm GS} _{+} ^{N -1 } }$ is the ground state of the Hamiltonian $H_+^{N-1}$~\eqref{eq:H+/-}, supported on $N-1$ sites, $H_+^{N-1}  \ket{\mathrm{GS}_+^{N-1}}=E_+^0  \ket{\mathrm{GS}_+^{N-1}}$.
We want to show that, in the localized phase, $\ket{ \Psi_0 (L; N) }$ is close to the ground state of $H^L$ in Eq.~\eqref{eq:QEast_block}, supported on $L$ sites.
In appendix~\ref{appdx:comp_var}, we demonstrate that the only contribution to the energy variance comes from the boundary term between $\ket{ {\rm GS} _{+} ^{N -1 } }$ and the string of empty sites.
By using $H_+^{N-1} - H_-^{N-1} = - e ^{-s} n_{N-1}$ from Eq.~\eqref{eq:H+/-}, it can be easily seen that neither the mean value, nor the variance of the energy evaluated in $\ket{ \Psi_0 (L;N) }$ depend on $L$, and they take the simple form
\begin{align}
  \braket{H^L}_{\Psi_0} & = E^0_{+} + \frac{1}{2} e^{-s} \delta, \label{eq:mean} \\
  \braket{ \Delta H^L }_{\Psi_0} & =  \frac{e^{-2s}}{2} \left[ \delta - \frac{1}{2} \delta^2 \right], \label{eq:mean_var}
\end{align}
where we have defined
$\delta=\bra{\mathrm{GS}_+^{N-1}} n_{N-1}\ket{\mathrm{GS}_+^{N-1}}$.
Eqs.~\eqref{eq:mean}, \eqref{eq:mean_var} show that both the mean energy and the variance of the state $\ket{ \Psi_0 (L; N) }$ (supported on $L$ sites) can be estimated from the knowledge of $\ket{ {\rm GS}_+ ^{N-1} }$ (supported on $N-1<L$ sites).
For small values of $\delta$, namely when the last spin of $\ket{ {\rm GS} _{+} ^{N-1} }$ is close to $\ket{0}$, the state $\ket{\Psi_0 (L;N)}$ is close to an eigenstate of $H^{L}$ for any $L$.
As can be seen in Fig.~\ref{fig:loc_GS}, this is precisely the case when $s>0$.
Eq.~\eqref{eq:mean_var} also shows that the quantity $\delta$ fully quantifies the energy variance of the extended state.
Accordingly, as long as the variance is smaller that the gap (which is sizable already for small positive values of $s$ and for all system sizes), we expect that the state $\ket{ \Psi_0 (L;N) }$ approximates the ground state of the Hamiltonian, independently of $L$.

Notice that the form of Eqs.~\eqref{eq:mean}, ~\eqref{eq:mean_var} is also valid (with the $E_+^0$ in~\eqref{eq:mean} replaced by the appropriate energy) if the factor $\ket{\mathrm{GS}_+^{N-1}}$ in Eq.~\eqref{eq:approx_gs} is replaced by any other eigenstate of $H^{N-1}_+$.
In Sec.~\ref{sec:eigenstates} we will use $\delta$ as a figure of merit for quantifying the number of eigenstates that admit an extension as the one in Eq.~\eqref{eq:approx_gs}, with small variance.
We will show that, for positive values of $s$, the property above is shared by several eigenstates of the model and not only by the ground state.

\subsection{Excited states for large system sizes} \label{sec:ext_eigen}

As we have shown above, by combining the ground state of small systems and strings of empty sites, it is possible to approximate ground states for large system sizes.
The construction utilizes two particular ingredients: the localization properties of the ground state, and the fact that the Hamiltonian annihilates a string of empty sites.
In this section we will construct an ansatz for excited states based on similar ideas.
Suppose $\ket{\phi_{\epsilon}^M}$ is an excited state of $H^M$ in Eq.~\eqref{eq:QEast_block} supported on $M$ sites, such that $H^M\ket{\phi_{\epsilon}^M}=E_{\epsilon}^M\ket{\phi_{\epsilon}^M}$.
The state
\begin{equation}\label{eq:approx_ex_st}
  \ket{ \Psi_{\epsilon} (L; N) } =  \ket{{\rm GS}_{+}^{N-1}}\otimes\ket{0}\otimes\ket{1}\otimes\ket{\phi_{\epsilon}^M},
\end{equation}
(such that $L=N+M+1$) exhibits similar properties as the one defined in Eq.~\eqref{eq:approx_gs}.
More precisely, as in the previous case, the only contribution to the energy variance comes from the boundary term between the ground states and the empty site and is given by Eq.~\eqref{eq:mean_var}.
The corresponding expectation value of the energy is $E = E^0_+ + E^M_{\epsilon} + e^{-s}\delta/2$.

Notice that the states in Eq.~\eqref{eq:approx_ex_st} can be arbitrarily close to an eigenstate of $H^L$ in Eq.~\eqref{eq:QEast_block} as long as $\delta$ is small enough.
Since the typical energy gap between two neighboring eigenstates in the middle of the spectrum for a generic Hamiltonian supported on $L$ sites scales as $2^{-L}$, in order to provide accurate approximations, $\delta$ needs to decrease at least as fast.
As illustrated by Fig.~\ref{fig:super_exp_decay}, $\delta$ decays super-exponentially, $\delta \sim \exp(- N\log{N})$, which implies that $N\log N \gtrsim L$ will be enough to satisfy that condition.
For very large system sizes ($L \rightarrow \infty$) this can be achieved if the ground state occupies a fraction of the sites $N/L$ approaching zero.
Therefore, the fraction $M/L$ of sites that can be occupied by an excited state  approaches one as we increase the system size.
As $M$ becomes larger, the states $\ket{\phi_{\epsilon}^M}$ can reach higher energies leading to {\it any} finite energy density for the states $\ket{ \Psi_{\epsilon} (L; N) }$. 

For more generic models as the ones discussed in appendix~\ref{appdx:generalization_localization}, we proved that the energy variance decays exponentially. Hence we can construct non-thermal states as the ones discussed above up to some finite (albeit not arbitrarily high) energy density.

It is worth stressing that the approximate eigenstates $\ket{ \Psi_{\epsilon} (L; N) }$ are non-thermal and, as long as $M = \order(L)$, they are exponentially many in system size $L$.
More precisely, for any given $N$, there are $2^{L-(N+1)}$ states of that form: a fraction $2^{-(N+1)}$ of the total number of states in the Hilbert space.

\paragraph*{Exploiting the maximally excited state}
The construction we just described provides an explicit way of addressing excited states at large system sizes by using eigenstates from smaller sizes.
In general, nevertheless, states of the form~\eqref{eq:approx_ex_st} do not need to fulfill an area law of entanglement, even if the leftmost $N$ sites are always in a product state with respect to the rightmost $M+1$ sites of the system, because a highly excited eigenstate $\ket{\phi_{\epsilon}^M}$ may have volume law entanglement.
Thus, the description of $\ket{ \Psi_{\epsilon} (L; N) }$ may require exponential resources.
However, there is at least one interesting exception to this situation, when the excited state $\ket{\phi_{\epsilon}^M}$ corresponds to the maximally excited state of the Hamiltonian $H^{M}$ in Eq.~\eqref{eq:QEast_block}, or equivalently, the ground state of $-H^M$ which also admits a MPS approximation.

If we choose $\ket{\phi_{\epsilon}^M}$ in Eq.~\eqref{eq:approx_ex_st} to be the maximally excited state $\ket{\phi_{\rm max}^M}$, we obtain an area-law state
$\ket{ \Psi_{\rm max} (L; N) }$, with  energy $E = E_+^0 + E^M_{\rm max} + e^{-s}\delta/2$.
Since we expect $E^M_{\rm max} \sim \order(M)$, as long as $M = \order(L)$, the resulting $\ket{\Psi_{\rm max} (L; N)}$ has finite energy density.
Moreover, its energy variance is $\braket{ \Delta H^L } < \delta$, so that in the localized phase it can be made arbitrarily small by increasing $N$, and the construction can provide approximate eigenstates.

From the exact diagonalization results above we know that even for small system sizes $\delta$ quickly reaches machine precision at least exponentially fast in $N$.
This means that even for modest $N$ its value becomes negligible in the construction above.
This immediately suggests an efficient numerical algorithm to construct quasi-exact highly excited eigenstates for system sizes much larger than the ones allowed by exact diagonalization, since we can use variational MPS methods to find the ground states of $H^N$ and $-H^M$ for chains of several hundred sites with extremely good precision~\cite{Banuls2019}.

\begin{figure}
	\includegraphics[width=0.5\textwidth]{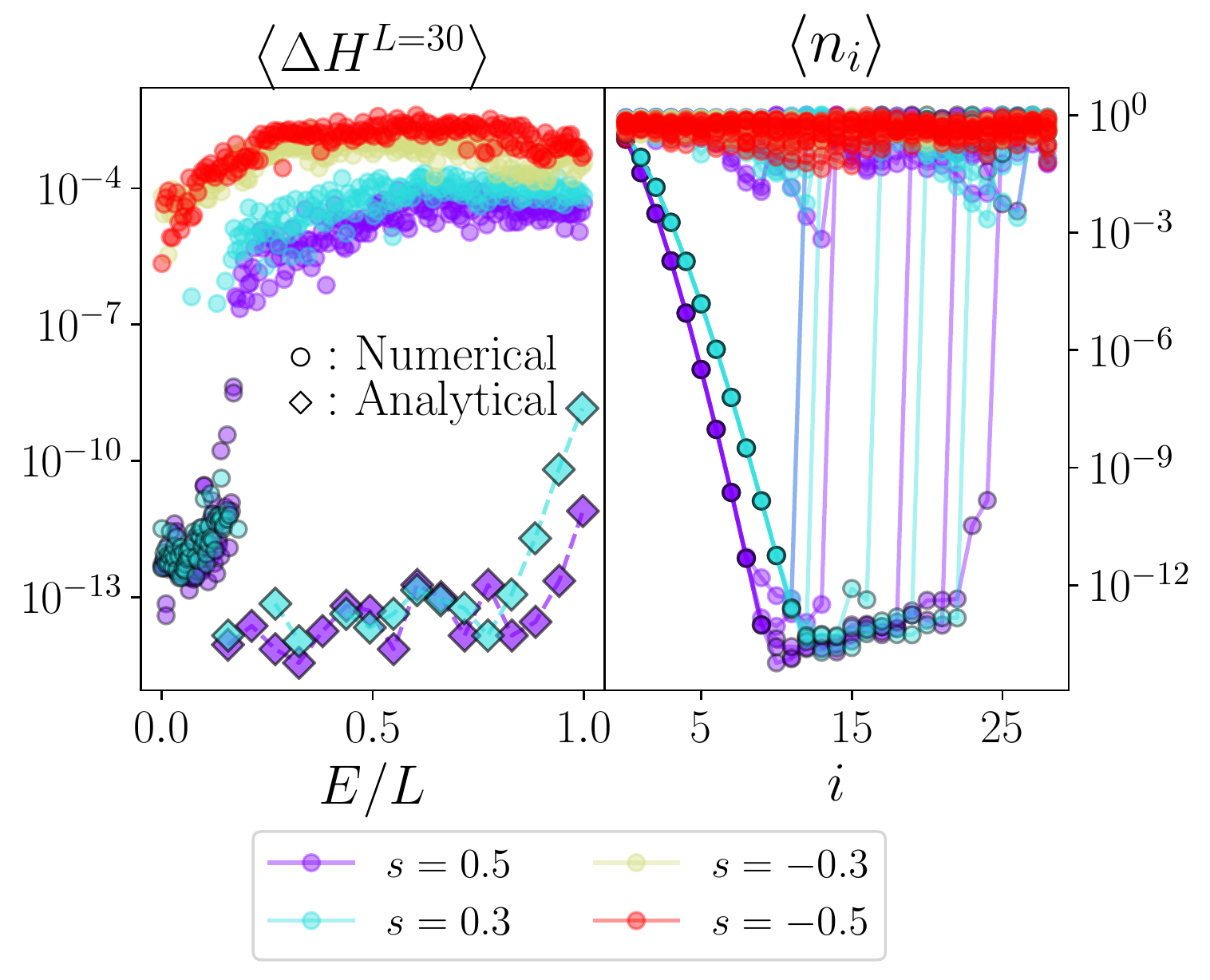}
	\caption{{\bf Energy variance(left) and single-site occupation expectation(right) of MPS approximations ($D=50$) to the excited states for $L=30$ sites.} (left) For small values of the energy densities and $s>0$, the MPS approximation are close to machine precision.
  A black edge indicates the MPS with variance below $10^{-8}$.
  For such states (right) we observe that the spatial distribution of the single site occupation corresponds to the profile of the analytical construction Eq.~\eqref{eq:approx_ex_st}.
	}
	\label{fig:mps_30}
\end{figure}

Fig.~\ref{fig:mps_30} illustrates the construction for a chain of size $L=30$.
In particular, we show the energy variance and occupation distribution of MPS approximations to excited states, found numerically as described in Sec.~\ref{sec:num_approx_var}.
For $s>0$ and small energy densities, for which the MPS provide almost exact eigenstates, we observe that their spatial profile indeed agrees with that of the analytical construction presented in this section.
Moreover, for $s>0$ the construction yields energy variances close to machine precision over practically the whole range of energies, where the direct MPS search is far from reaching an exact eigenstate.

\section{The super-spin picture}\label{sec:frankenstein}

Here we exploit the results from previous sections to engineer a large class of states with small variance.
The basic idea is concatenating several blocks of $N+2$ sites, each of them in  one of two mutually orthogonal states,
\begin{align}\label{eq:frank_building_blocks}
    \ket{ \Tilde{0} } := \ket{0}^{\otimes{N+2}},\quad &\ket{ \Tilde{1} } := \ket{ 1} \otimes \ket{{\rm GS}^{N}_{+} } \otimes \ket{ 0 } .
\end{align}
We identify the subspace spanned by these two vectors with the Hilbert space of a \emph{super-spin}.
For a system of size $L$, we can thus construct $2^{L/(N+2)}$ orthogonal states
$\ket{\tilde{s}} \in \{ \ket{\Tilde{1}}, \ket{\Tilde{0}} \}^{\otimes L/(N+2)}$.
All such states fulfill an area law and can be approximated as a MPS insofar
as $\ket{{\rm GS}^{N}_{+}}$ does.

The states in this set retain long memory of their initial conditions and stay weakly entangled under time evolution, as we will see in the following subsections.
The key dynamical property that we exploit is their energy variance, which can be easily computed using the same procedure as in section~\ref{subsec:exact_GS}.
Since $\ket{ \Tilde{0} }$ blocks do not contribute to the variance,  the only contributions come from blocks in $\ket{ \Tilde{1}}$, and correspond to the value computed in Eq.~\eqref{eq:mean_var}
\begin{equation}\label{eq:var_frank}
  \braket{ \Delta H } _{\tilde{s}} = \sum_{k=1}^{\mathcal{M}} \braket{ \Delta H_k }_{\Psi_0^N}
  = \mathcal{M} \braket{ \Delta H^N }_{\Psi_0^N} < \mathcal{M} \delta ,
\end{equation}
where the index $k$ runs over the positions of the occupied super-spins and $\mathcal{M}$ is the Hamming weight of $\ket{\tilde{s}}$.
It is important to stress that $\mathcal{M}$ is potentially unbounded in the thermodynamic limit, in which case the variance becomes unavoidably large.

The energy can also be easily computed,
\begin{equation} \label{eq:en_frank}
  \braket{  H }_{\tilde{s}} = \mathcal{M} (E^0_{+} + e^{-s}\delta /2 ) .
\end{equation}
Equations \eqref{eq:var_frank} and \eqref{eq:en_frank} show that if $\mathcal{M}$ is chosen appropriately we can construct states with high energy and exponentially small variance in $N$.
Notice however that, as we want states with small energy variance, we need to introduce limitations on the values of $\mathcal{M}$.

From Eq.~\eqref{eq:var_frank}, note that the variance of the super spins cannot exceed the value $\delta L/(N+2) \lesssim 2^{-N\log(N)} L/(N+2) < 2^{-N} L/(N+2)$, since for any given super spin we can accommodate at most $\mathcal{M} = L/(N+2)$ occupied blocks.
Clearly, we have the freedom of choosing $N$ at will.
However, the choice will affect the variance of the super spins, and the dimension of the Hilbert space spanned by them.
It is illustrative to mention a few interesting cases.
({\it i}) If $N \sim \log L^{\beta}$ with $\beta > 1$ then, the state with $\mathcal{M}$ occupied super spins can have a large variance, exponentially larger than $2^{-L}$: $\braket{ \Delta  H }_{\tilde{s}} < L^{1-\beta}/\log L^{\beta}$.
The dimension of the corresponding Hilbert space is of the order $2^{L/\log L^{\beta}}$.
({\it ii}) An opposite scenario is when $N \sim L/\log L$.
In this case, the variance is small $\braket{ \Delta  H }_{\tilde{s}} < 2^{-L/\log L} \log L $, but the dimension of the Hilbert space is linear in $L$.
({\it iii}) An interesting intermediate example consists in $N \sim L^{\alpha}$, with $0 < \alpha < 1$.
Here the variance is $\braket{ \Delta H }_{\tilde{s}} < L^{1-\alpha} 2^{- L^{\alpha}}$ and the dimension of the Hilbert space scales as $2^{L^{1-\alpha}}$, which is sub-exponential in system size.
In the following section we will show how the variance of an initial state can be use to quantify its slowness.
The super-spin picture provides a flexible platform where one can choose the appropriate trade-off between the dimension of the Hilbert space and the dynamical activity of the super-spin vectors that span it.

\subsection{Dynamical properties of the super-spin states}

The memory of the initial state during time evolution admits a general bound based on the initial energy variance.
For an initial state $\ket{\psi_0}$, we define the overlap~\cite{Kim2015}
\begin{equation}\label{eq:a_t}
  a (t) = \left| \braket{ \psi_0 | \psi_t } \right|^2 = \Tr \left( \rho_0 \rho_t \right),
\end{equation}
where $\ket{\psi_t}$ is the state at time $t$ and  $\rho_{t} = \ket{ \psi_{t} } \bra{ \psi_{t} }$ the corresponding density matrix.
Using the Cauchy-Schwarz inequality,
\begin{equation}\label{eq:d2aM_bound}
  \begin{split}
    \left| \frac{d^2 a (t)}{dt^2} \right|  \leq & \| \left[\rho_0, H\right] \|_F \| \left[\rho_t, H\right] \|_F \\
    = & \sqrt{4 \braket{ \Delta H }_{\psi_0} \braket{ \Delta H }_{\psi_{t}}} = 2 \braket{ \Delta H }_{\psi_{0}} ,
  \end{split}
\end{equation}
where $\|\cdot\|_F$ denotes the Frobenius norm.
In the second line we used the fact that for the commutator with the Hamiltonian this norm does not depend on time.
Exploiting Eq.~\eqref{eq:d2aM_bound} we can compute the memory of the initial state as
\begin{equation}
  \begin{split}
    \left| a(t)-a(0) \right| = & \left| \int_0^t \int_0^{\tau} \frac{d^2 a (\tau')}{d\tau'^2} d\tau d\tau' \right| \\
    \leq & 2 \braket{ \Delta H }_{\psi_{0}}  \int_0^t \int_0^{\tau} d\tau d\tau' ,
  \end{split}
\end{equation}
which leads to the bound
\begin{equation}\label{eq:bound_overlap}
  a (t) \geq 1 - \braket{ \Delta H }_{\psi_0} t^2,
\end{equation}
where we used $a(0)=1$, $\frac{da}{dt}|_{t=0}=0$ and $a(t) \leq a(0)$.
Eq.~\eqref{eq:bound_overlap} is a general bound on the memory of a time evolved state based on the energy variance of the corresponding initial state.

The bound in Eq.~\eqref{eq:bound_overlap} can be used to bound the growth of the entanglement entropy of an arbitrary subsystem.
According to the Fannes inequality, for any pair of density matrices $M_1$, $M_2$, of dimensions $\mathcal{D}\times\mathcal{D}$~\cite{Nielsen:2011:QCQ:1972505},
\begin{equation}\label{eq:fannes_audenaert}
  \left| S(M_1) - S(M_2) \right| \leq 2T \log(\mathcal{D}) - 2T\log(2T),
\end{equation}
where $T=\frac{1}{2}\|M_1 - M_2 \|_1=\frac{1}{2}\Tr[\sqrt{(M_1 - M_2)(M_1 - M_2)^{\dag}}]$ is the trace distance between both matrices, and $S(M) = - \Tr \left[ M \log(M) \right]$ is the von Neumann entropy.
We can apply Eq.~\eqref{eq:fannes_audenaert} to the reduced density matrix of a subsystem at the initial time and after evolution~\footnote{Note that Eq.~\ref{eq:fannes_audenaert} holds if $T \leq 1/2e$.
This does not spoil the results at larger times, since we can use the weaker relation $\Delta S \leq 2T \log\mathcal{D} + 1/(e \log 2)$, which qualitatively gives the same scaling.}.
Given some partition $\mathcal{H} = \mathcal{H}_A \otimes \mathcal{H}_B$ of the Hilbert space, let us define the (time-dependent) reduced density matrix $\sigma_{\tau} = \Tr_B (\rho_{\tau})$.
Contractivity of the trace norm ensures
\begin{equation}\label{eq:contractivity}
  \frac{1}{2} \| \sigma_t - \sigma_0 \|_1 \leq \frac{1}{2} \| \rho_t - \rho_0 \|_1 \leq t\sqrt{ \braket{ \Delta H }_{\psi_0}},
\end{equation}
where in the second inequality we used
\begin{equation}
  \frac{1}{2}\| \rho_t - \rho_0 \|_1 = \sqrt{1- \left| \braket{ \psi_0 | \psi_t } \right|^2} = \sqrt{1-a(t)} ,
\end{equation}
and Eq.~\eqref{eq:bound_overlap}.
Notice that Eq.~\eqref{eq:contractivity} sets an explicit bound on how fast the expectation value of any local observable can change when starting from a super-spin state.

By plugging Eq.~\eqref{eq:contractivity} in Eq.~\eqref{eq:fannes_audenaert}, we can bound the growth of the entanglement entropy as
\begin{equation}\label{eq:entropy_bound}
  \left| S(\sigma_t) - S(\sigma_0) \right| \leq 2 t \sqrt{ \braket{ \Delta H }_{\psi_0}}  \log \left(\frac{\mathcal{D}}{2 t \sqrt{ \braket{ \Delta H }_{\psi_0} } } \right) .
\end{equation}
Eq.~\eqref{eq:entropy_bound} provides a general bound on the growth of the entanglement entropy of a subsystem based on the energy variance of the initial extended pure state, and the dimension $\mathcal{D}$ of the subsystem. 

The bounds on the memory of the initial state in Eq.~\eqref{eq:bound_overlap} and the growth of the entropy in Eq.~\eqref{eq:entropy_bound}, can be straightforwardly applied to the super-spins $\ket{\tilde{s}}$, defined in Sec.~\ref{sec:frankenstein}.
In the particular case when $\rho_0 = \ket{\tilde{s}} \bra{\tilde{s}}$ (supported on $L$ sites) we can bound the memory of the initial conditions by using Eq.\eqref{eq:bound_overlap} and $\braket{ \Delta H }_{\tilde{s}} < \mathcal{M} \delta$.
Namely,
\begin{equation}\label{eq:memory_bound_phi_M}
  | \langle \tilde{s} (t) | \tilde{s} (0) \rangle |^2 \geq 1 - t^2 \mathcal{M} \delta .
\end{equation}
Accordingly, if we take $\sigma_0$ to be
the corresponding reduced density matrix for a region of $N \ll L$ sites, $\mathcal{D} = 2^{N}$.
The bound in Eq.~\eqref{eq:entropy_bound} then reads
\begin{equation}\label{eq:entropy_bound_phi_M}
  \begin{split}
    \left| S(\sigma_t) - S(\sigma_0) \right| \leq & 2t \sqrt{\mathcal{M} \delta} \left( N  \log \left( 2 \right) - \log \left( 2t \sqrt{\mathcal{M} \delta} \right)\right).
  \end{split}
\end{equation}

In the previous sections we showed that in the localized region $\delta$ decreases exponentially with $N$.
As a consequence, if $\mathcal{M}$ is sufficiently small, Eq.~\eqref{eq:memory_bound_phi_M} and Eq.~\eqref{eq:entropy_bound_phi_M} provide strong bounds on the dynamics of $\ket{\tilde{s}}$.
Specifically, in order to erase half of the memory of the initial state, i.e. $| \langle\tilde{s} (t^*) | \tilde{s} (0) \rangle |^2 \leq 1/2$, the dynamics needs at least exponentially long times in $N$, $t^* \gtrsim (\mathcal{M} \delta /2)^{-1/2} $.
At the same time, for entangling a sub-region of size $N$, i.e. $2 t \sqrt{\mathcal{M} \delta} N  \sim 1$, the time evolution necessitates exponential times of the order  $t^* \sim ( 2 N \sqrt{\mathcal{M}\delta})^{-1}$.
We conclude that the dynamics of the states $\ket{\tilde{s}}$, in order to entangle a sub-region, requires exponentially long time in the subsystem size. 

The states $\ket{\tilde{s}}$ can then be seen as an orthonormal set of quasi-conserved area-law vectors, and any superposition of them will result in a state whose dynamics at short times is governed by dephasing only.
The super-spin picture thus provides an effective description of a subset of the Hilbert space in the thermodynamic limit which evolves slowly in time, is weakly entangled, and efficiently simulable.

The results in Ref.~\cite{Horssen2015} can be reinterpreted from a super-spin point of view.
It was numerically argued there, for the case of periodic boundary conditions, that for certain product states the dynamics exhibits a slow growth of the entanglement entropy, exponential in system size.
The {\it slowness} of the state was quantified by the number of empty sites following an occupied one.
Since the previous statements about the energy variance of $\ket{\tilde{s}}$ do not depend on the boundary conditions and, as argued in section~\ref{sec:ground_state},
the block ground state for $s>0$ is very close to the product state $\ket{ 1000\dots }$, the bound in Eq.~\eqref{eq:entropy_bound_phi_M} gives a rigorous interpretation of the previous numerical observations.

\paragraph*{Extensions}
The super-spin construction described above can be made more general in several ways.
On the one hand, a larger set of states can be constructed by allowing not only the ground state, but also (sufficiently localized) excited states as building blocks $\ket{\phi_{\epsilon}^M}$.
In Sec.~\ref{sec:eigenstates} we show that such excited states do actually exist.
On the other hand, by combining the super-spin picture with the excited state construction in Sec.~\ref{sec:ext_eigen} we can also construct states with finite energy density.
Namely, we can construct states $\ket{\mathcal{S} }= \ket{\tilde{s}} \otimes \ket{\phi_{\rm max}^M}$ with energy $\braket{  H }_{\mathcal{S}} = \mathcal{M} (E^0_{+} + e^{-s}\delta /2 ) + E^M_{\rm max}$ and energy variance $\braket{  \Delta H }_{\mathcal{S}} < \mathcal{M} \delta$.
By increasing $M$ the energy density can be increased, but at the cost of reducing the dimension of the super-spin subspace to  $2^{(L-M)/(N+2)}$.

\section{Non-thermal Excited states in small system sizes}\label{sec:eigenstates}

\begin{figure}
  \includegraphics[width=0.5\textwidth]{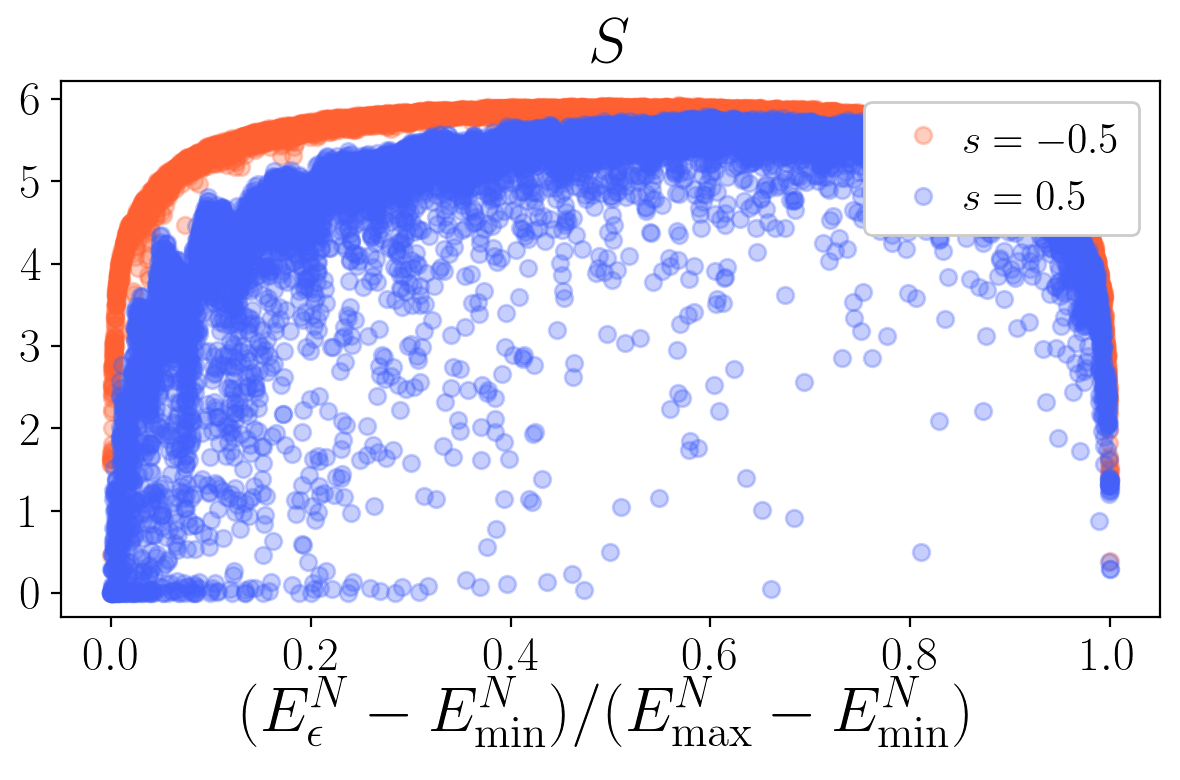}
  \caption{{\bf Entanglement Entropy of the eigenstates vs. energy density, $N=14$.} In the two regimes of positive and negative $s$, the entanglement entropy of the eigenstates in the middle of the chain shows abrupt changes from {\it normal} eigenstates at $s=-0.5$, to {\it anomalous} eigenstates, where many of them have small entropy in the middle of the spectrum at $s=0.5$.}
  \label{fig:ent_vs_ener}
\end{figure}

In the following we explore the properties of the whole Hamiltonian spectrum using exact diagonalization for small system sizes.
The results indicate a substantial change in the properties of eigenstates across the spectrum in the region $s>0$.
In particular, in this region many localized eigenstates can be found, beyond the ground state, which can be used in the constructions of the previous sections.

\subsection{Entropy of the exact eigenstates}

Since eigenstates of the Hamiltonian incorporate the whole information about the dynamics of the system, their entanglement entropy is often used as an indicator of the associated dynamical behavior.
In Fig.~\ref{fig:ent_vs_ener}, we show the entanglement entropy in the middle of the chain of $N = 14$ spins from exact diagonalization, for two values of $s$.
For negative $s$, the entanglement entropy of eigenstates exhibits behavior compatible with a thermalizing system --- apart from the extremes of the spectrum, most of the eigenstates have large entanglement, almost saturating the upper bound given by system size.
In contrast, for positive values of $s$ a considerable number of excited eigenstates have low entanglement.
This is an indication of non-thermal eigenstates and reminiscent of the quantum scars found in the PXP model \cite{Turner2018}, but here we observe this behavior for a much larger number of states.

\begin{figure}
  \includegraphics[width=0.52\textwidth]{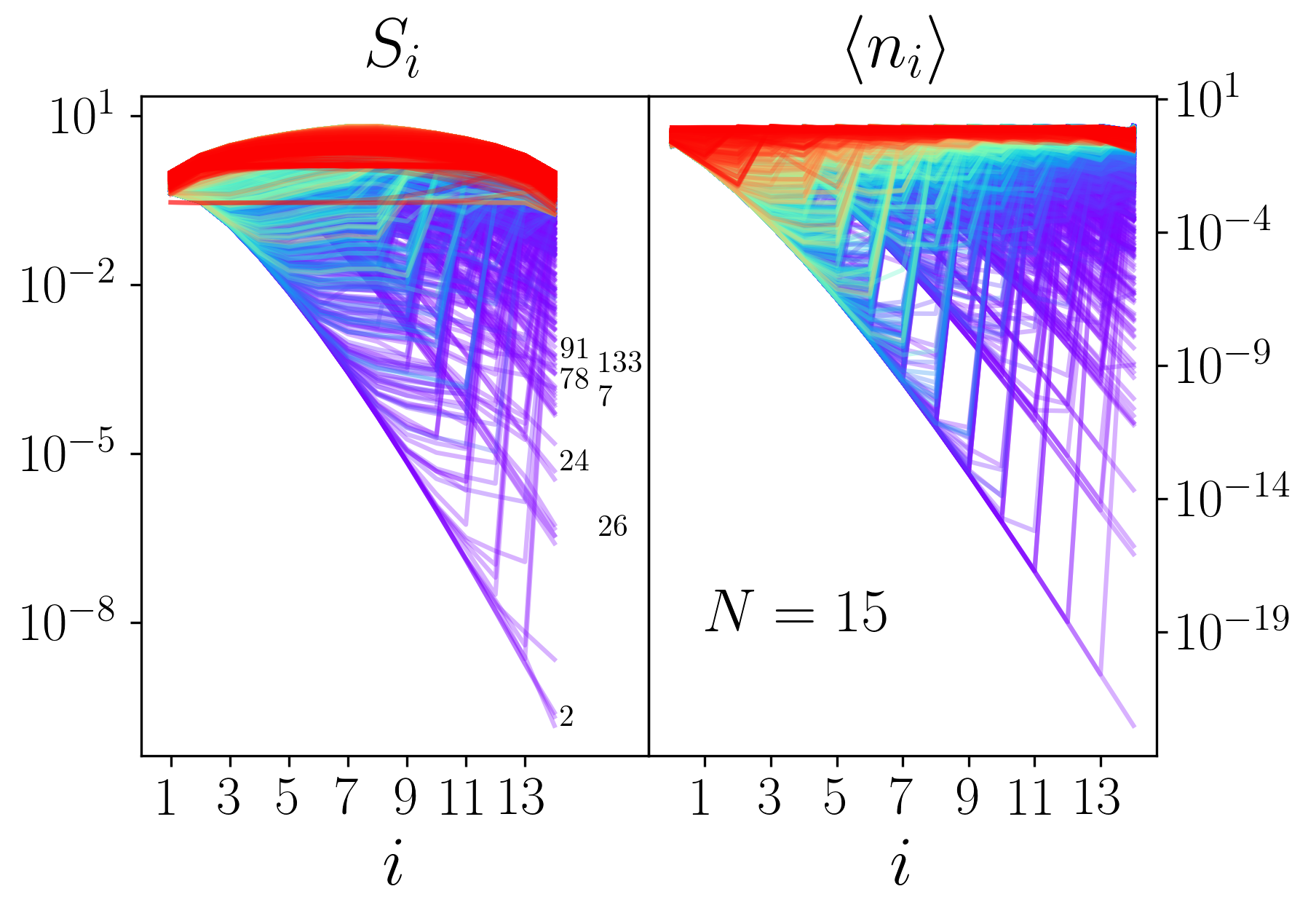}
  \caption{{\bf  Entanglement entropy and local occupation as a function of the position on the chain.} For several eigenstates both quantities decay exponentially with the distance to the left edge.
  The labels in the entropy plot indicate the indices of (some of) the eigenvectors ordered by increasing energy.
  Colors indicate the energy of the eigenstates from purple (low energy) to red (high energy).
  The figures show data for $s=0.5$ and $N=15$.
  }
  \label{fig:ent_support}
\end{figure}

In order to collect detailed information about the distribution of the entanglement along the spin chain, we compute, for each eigenstate, the entanglement entropy with respect to all possible cuts of the chain, $S_i= S\left( \rho_i \right)$, where $\rho_i$ is the reduced density matrix obtained when tracing out all but the leftmost $i$ spins.
In Fig.~\ref{fig:ent_support}, we plot the entanglement entropy $S_i$ and single site occupation $\braket{n_i}$ as a function of the position of the cut (respectively the site) $i$ for all eigenstates in the case $s=0.5$ and $N=15$.
The figure suggests a peculiar {\em heterogeneous entanglement} structure of a significant number of eigenstates, for which both quantities decay exponentially as the cut moves to the right.
In other words, for many eigenstates, the spins far from the left edge are almost in a product state with the rest of the system, and the corresponding sites are almost empty.
These results are qualitatively similar to the ones discussed in Sec.~\ref{sec:ground_state} where we analyzed the localization properties of the ground state.

\begin{figure}
  \includegraphics[width=0.5\textwidth]{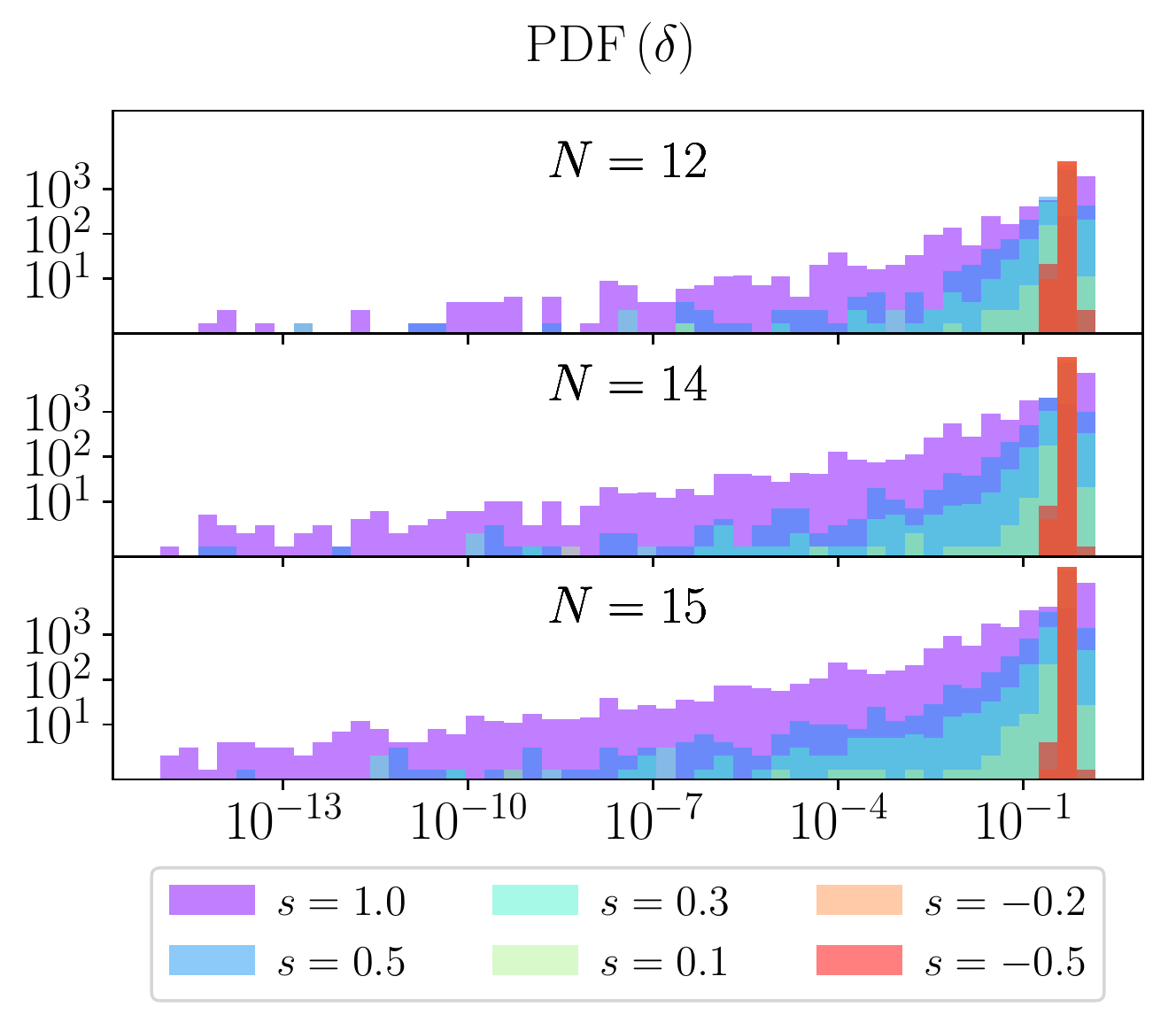}
  \caption{{\bf Probability density function of $\delta$, log-log scale.} For any given eigenstate we compute the expectation value of the projector onto occupied site $n_{N}$.
  As in the previous cases, the distribution is strongly peaked for negative values of $s$, and it develops very long tails for positive values.
  }
  \label{fig:P_nl}
\end{figure}

\subsection{Small-$\delta$ eigenstates}

We diagonalize the Hamiltonian $H_+^N$ in Eq.~\eqref{eq:H+/-} for different system sizes $N$ and values of $s$.
Given the set of eigenvectors, we consider the probability distribution of the last site occupation $\delta = \braket{n_N}$, the parameter which, as discussed in Sec.~\ref{subsec:exact_GS}, quantifies the variance of the extended states $\ket{ \phi_{\epsilon}^N } \otimes \ket{ 0 \ldots 0 }$.
Fig.~\ref{fig:P_nl} shows the histogram of the corresponding probability density function ${\rm PDF} ( \delta )$.
Notice that, for positive values of $s$, many eigenstates exhibit surprisingly small values of $\delta$.
Namely, there are several eigenstates $\ket{ \phi_{\epsilon}^N }$ such that the energy variance of the state $\ket{ \phi_{\epsilon}^N } \otimes \ket{ 0 \ldots 0 }$ can be bounded by extremely small values.

\begin{figure}
  \includegraphics[width=0.5\textwidth]{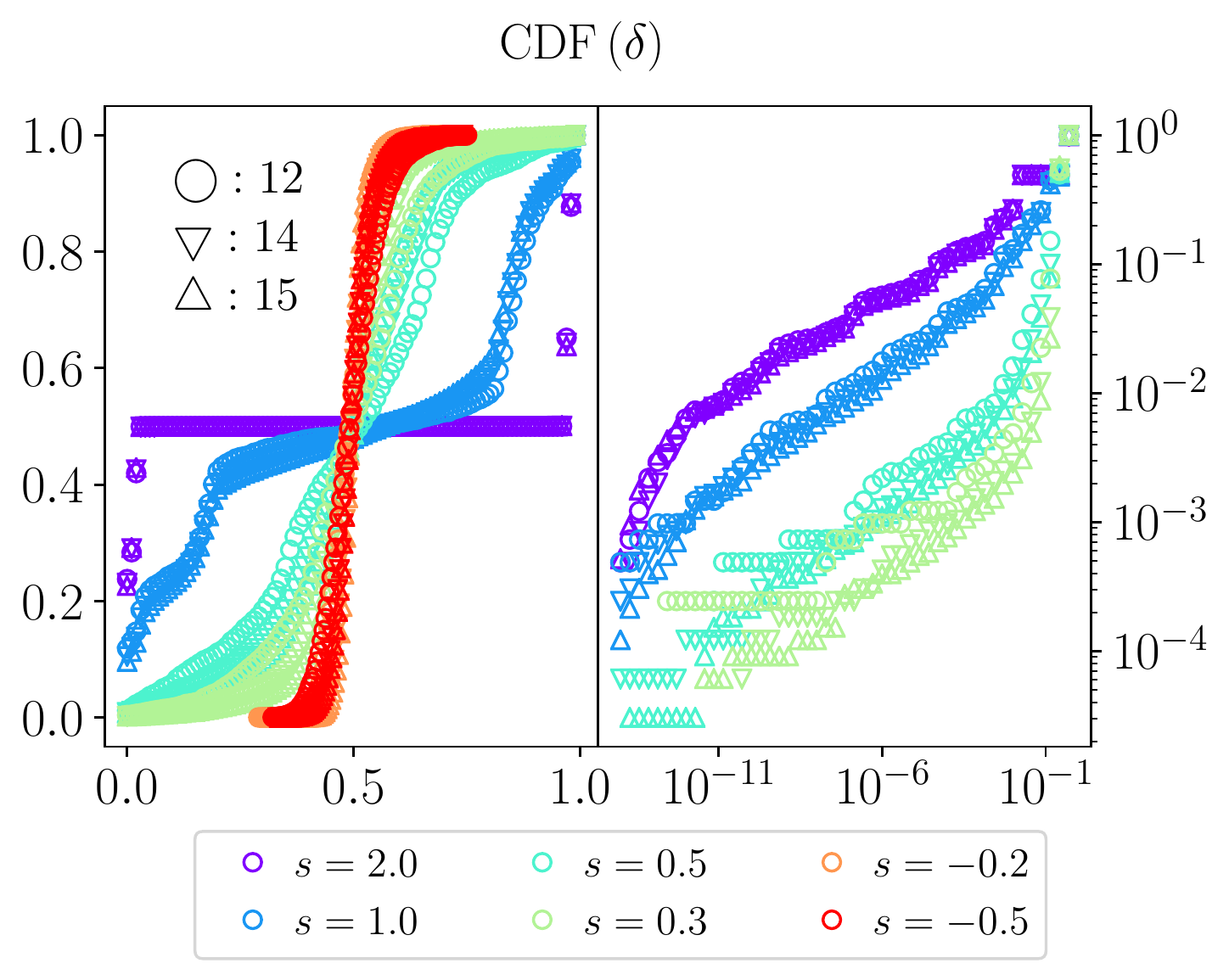}
  \caption{{\bf Cumulative distribution function of $\delta$.} Same data as in Fig.~\ref{fig:P_nl}.
  The CDF shows a steep curve for negative values of $s$, where most of the eigenstates have large values of $\braket{ n_{N} }$.
  When $s$ is positive the CDF shows fat tails extending to values close to machine precision.
  For large enough values of $s$, the fraction of eigenstates with small $\braket{ n_{N} }$ seems to be size independent.
  }
  \label{fig:CDF_nl}
\end{figure}

\begin{figure}
  \includegraphics[width=0.5\textwidth]{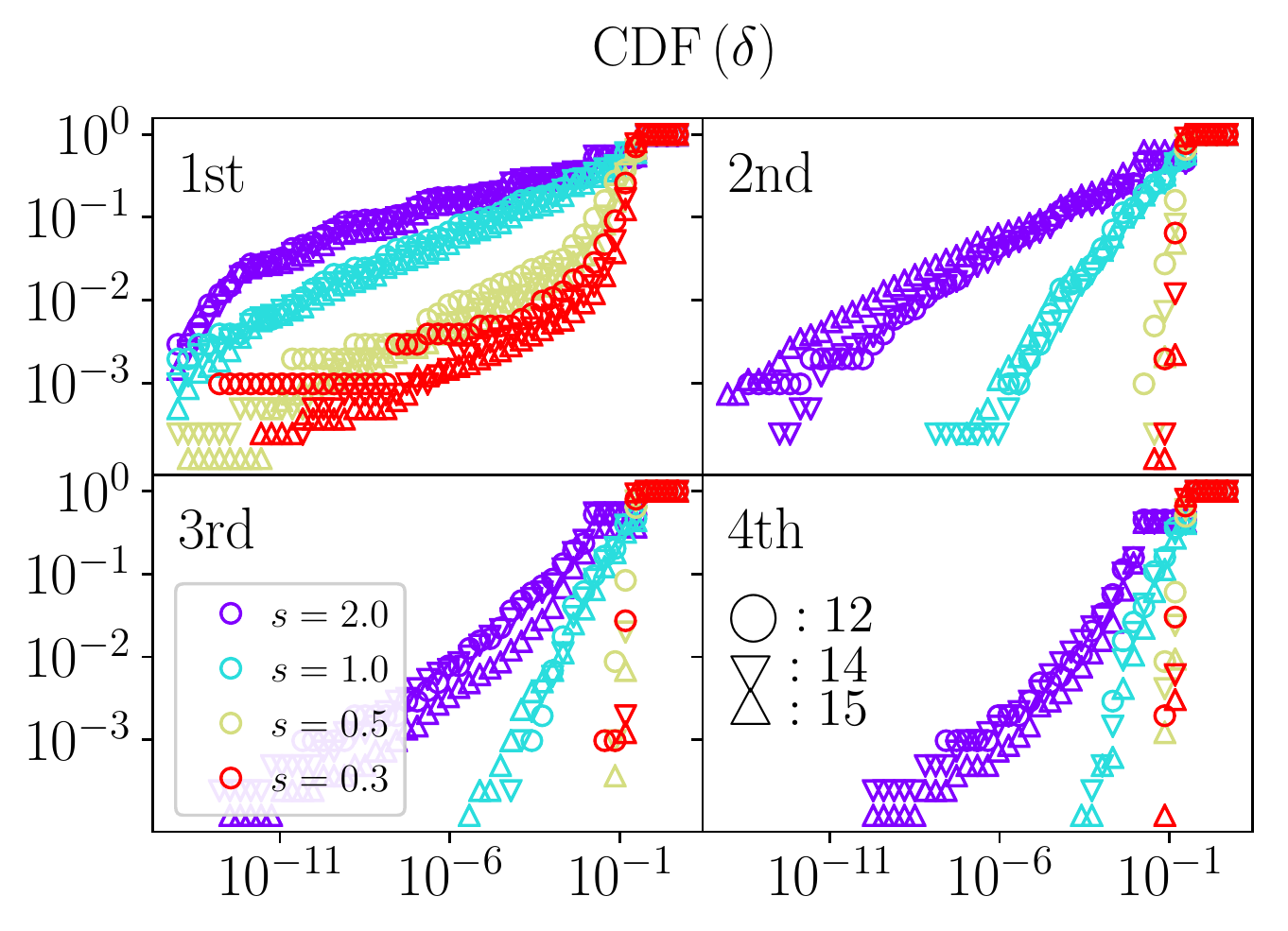}
  \caption{{\bf Energy-resolved cumulative distribution function of $\delta$.} We split the eigenstates in four equal-size energy intervals, ordered by growing energy.
  Most of the  eigenstates with small $\delta$ are concentrated at low energy.
  For the largest $s=2$, the count for eigenstates with small $\delta$ increases at all energies.
  }
  \label{fig:log_q4_CDF_nl}
\end{figure}

In order to quantify the number of eigenstates with small $\delta$, in Fig.~\ref{fig:CDF_nl} and \ref{fig:log_q4_CDF_nl} we consider the cumulative distribution function ${\rm CDF} (\delta)$.
In particular, in Fig.~\ref{fig:CDF_nl} we observe an abrupt change from negative values of $s$, where most of the eigenstates have large values of $\delta$ to positive $s$, where more and more eigenstates have very small values.
For the sizes accessible by exact diagonalization, the fraction of eigenstates with small $\delta$ does not seem to depend on the size of the system.
In Fig.~\ref{fig:log_q4_CDF_nl} we show the energy-resolved CDF.
In particular, we divide the spectrum in four intervals of equal energy width, which we number in order of increasing energy.
The figure shows that most of the small-$\delta$ eigenstates are concentrated in the lower part of the spectrum, in agreement with the results in Fig.~\ref{fig:ent_support}.
As $s$ increases, we observe that the number of eigenstates with small $\delta$ values grows for all energy regions, as we indeed expect from the discussion in the previous sections and the smaller localization length.

\begin{figure}
  \includegraphics[width=0.5\textwidth]{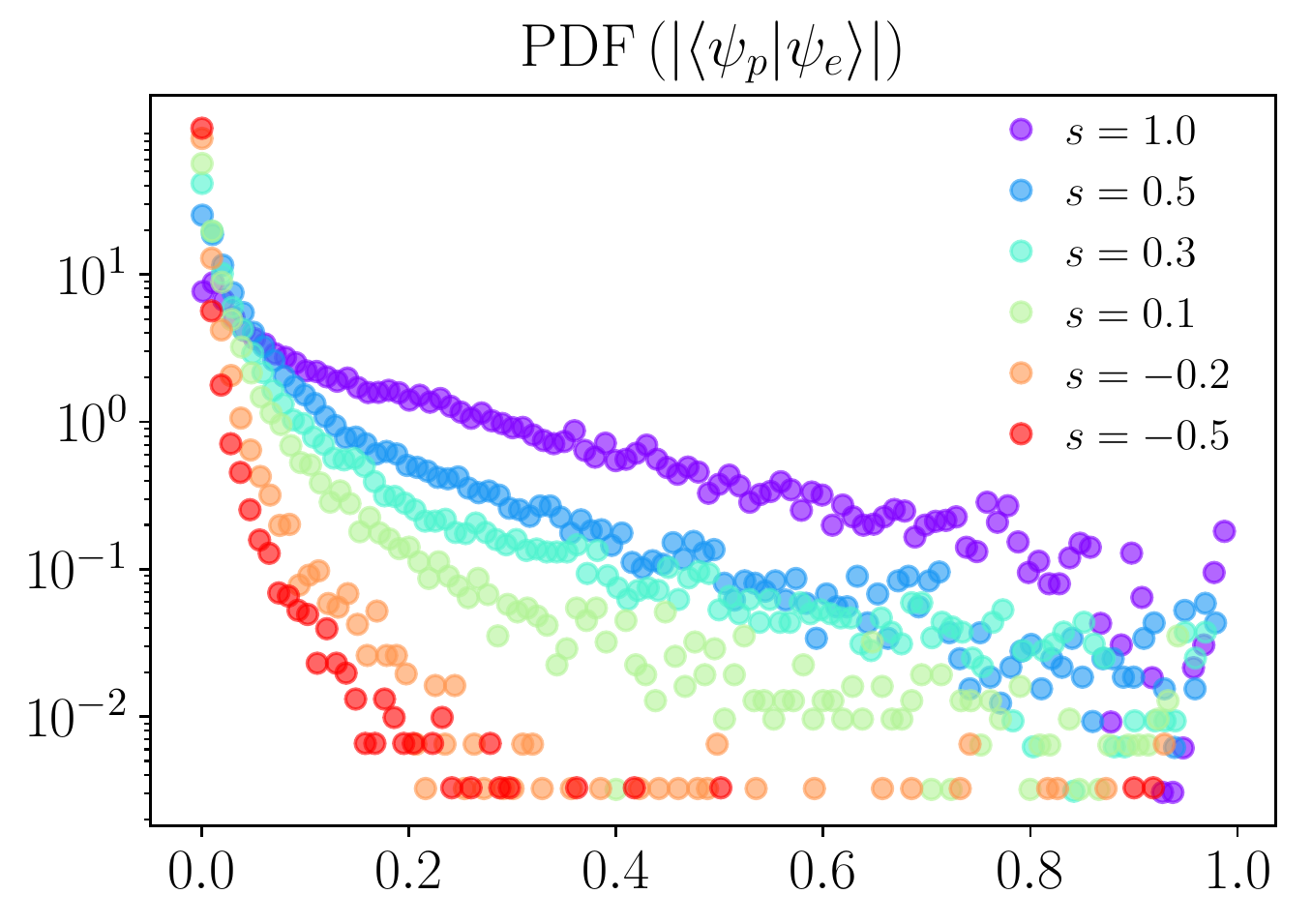}
  \caption{{\bf Probability density function of the estimated maximal overlap with a product state.}
  For negative values of $s$ most of the eigenstates have small overlap with product states.
  For positive values of $s$ the distribution develops long tails: many eigenstates have considerably large overlap with product states.
  Data is shown for $N=15$.
  }
  \label{fig:P_overlap}
\end{figure}

\subsection{Geometric entanglement}\label{sec:prod_state}

The geometric entanglement of a state, defined as its minimum distance to a product state, also provides  interesting insights about the properties of the eigenstates.
Given a pure state, the geometric entanglement can be found by maximizing its overlap with a product state.
Although it is possible to solve this optimization problem with exact or approximate numerical algorithms, in our case this is unpractical,
since we need to repeat the calculation for each eigenstate.
Instead, we apply a simpler \emph{one-sweep} truncation strategy to construct an approximation to the closest product state.
Namely, for each eigenstate, we sequentially perform a singular value decomposition with respect to each cut of the chain and keep only the largest singular value for each of them.
The resulting product state, once normalized, provides a lower bound to the maximum overlap.

In Fig.~\ref{fig:P_overlap} we plot the probability density function (over all energy eigenstates) of this estimate for the maximum overlap.
For negative values of $s$, most of the eigenstates have a small overlap with product states (as expected for an ergodic system).
For positive values of $s$, the distributions develop a fat tail towards small values of $\delta$, indicating that many eigenstates have a large overlap with product states.

\begin{figure}
	\includegraphics[width=0.5\textwidth]{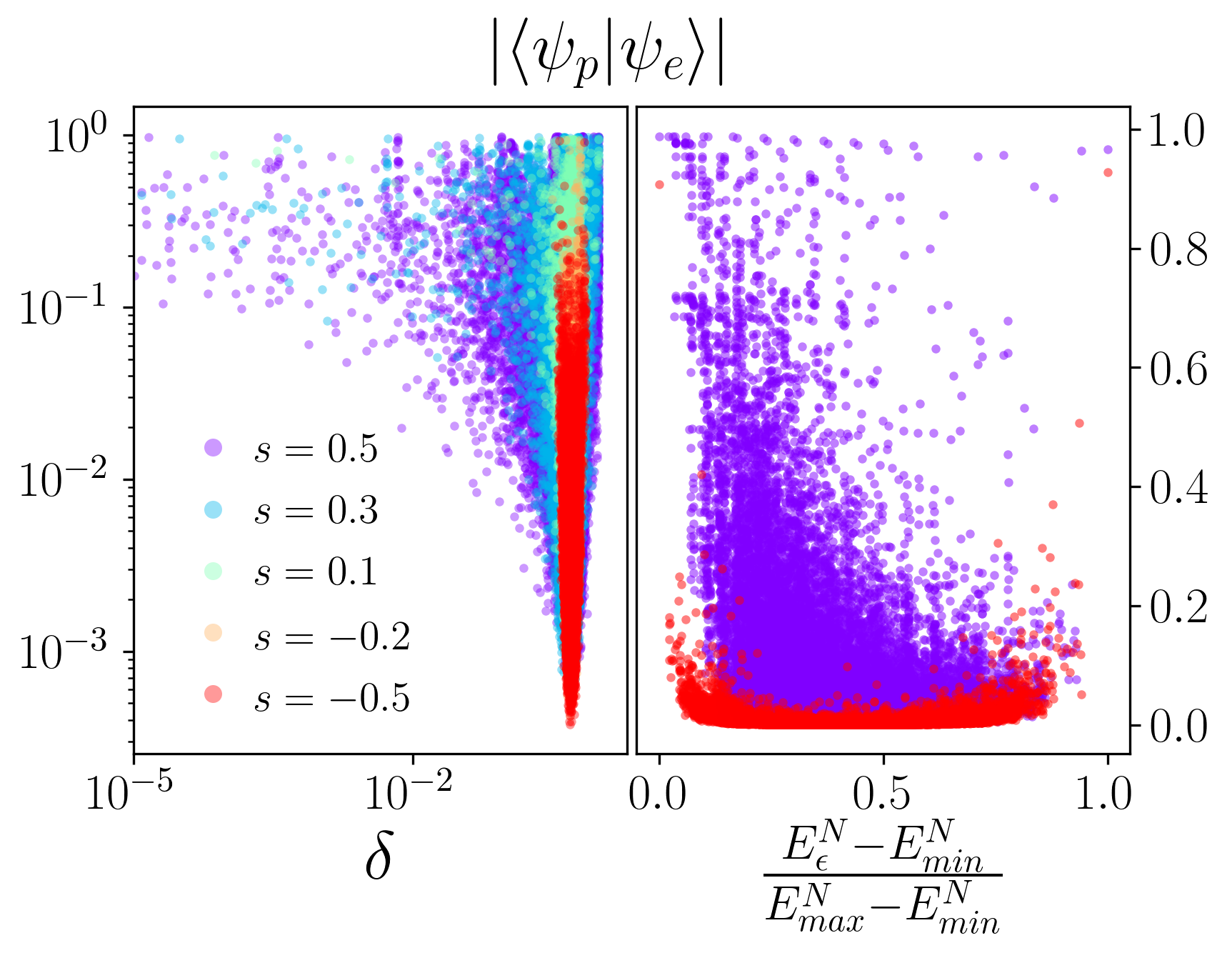}
	\caption{{\bf Scatter  plot of the overlap with product states vs. $\delta$ (left), and energy density (right).}
	The states with large overlaps are mostly concentrated at small energy densities, although large values can be found everywhere along the spectrum.
  We show data for $N=15$.
  On the right, only data points for $s= \pm 0.5$ are shown.
  }
	\label{fig:scatter_overlap_delta}
\end{figure}

An alternative view of this feature is demonstrated in Fig.~\ref{fig:scatter_overlap_delta}, which shows the value of the overlap for each eigenstate, as a function of the corresponding $\delta$ (left) or energy density (right).
Small values of $\delta$ are strongly correlated with large --- order $\order (1)$ --- overlaps with product states.
They are mostly concentrated at small energy densities, but Fig.~\ref{fig:scatter_overlap_delta} shows that large overlaps can actually be found at arbitrary energy densities.

\subsection{Numerical approximation of non-thermal excited states for large sizes}\label{sec:num_approx_var}

The discussion in Sec.~\ref{sec:ext_eigen} indicates the existence of highly excited states with small entanglement, that can be written as MPS.
We can thus try to find them with numerical methods.
There are several possible variations of DMRG to try and target excited states~\cite{Yu2017}.
The simplest one attempts to find the MPS that minimizes the expectation value of the operator $W=(H-\lambda)^2$, where $\lambda$ is the target energy of the desired states.
Since $H$ is a matrix product operator, also $W$ has that form and the minimization can be run efficiently with standard MPS algorithms.
We use this tool to probe the whole energy spectrum for eigenstates that can be approximated by MPS.

In the numerical study we fix the system size to $L=30$ and the bond dimension to $D=50$.
For several values of $s$, we then collect $300$ data points, uniformly distributed in energy (excluding the edges of the spectrum).
In Fig.~\ref{fig:mps_30} (left) we show the energy variance as a function of the energy density.
We observe that, for $s>0$ and low energy, the algorithm produces MPS with variance close to machine precision.
Fig.~\ref{fig:mps_30} (right) shows the profile of the expectation value of the single-site occupation number $\braket{n_i}$ as a function of the site $i$.
At low energy densities, and positive $s$, the optimization finds states with a structure that resembles $\ket{ \Psi_{\rm max} }$, with exponentially decreasing occupation from the left edge to the right until a certain site, where the occupation abruptly increases to stay close to one until the right edge.
In Fig.~\ref{fig:mps_30} we marked with a black circle the states with energy variance smaller than $10^{-8}$.
We find that all the states with small energy variance have an exponential tail which starts from the left edge.
According to our analytical construction in Sec.~\ref{sec:exact_eigs}, the position of the jump should correspond to the energy of the state.
For large energies such construction becomes harder to capture, and the optimization is forced to search for a trade-off between accurate target energy or small energy variance.
Notice that our optimization is not tailored to search for this specific construction, as each run starts from a random initial MPS.

\section{Discussion and Generalizations}\label{sec:generalization}

From the detailed study of the quantum East model, we have shown that for a broad class of constrained quantum Hamiltonians, there is a phase in which the (occupation of the) ground state is localized, and correspondingly slow dynamics arises. The quantum East model, specifically, is known~\cite{Garrahan2007,Banuls2019} to have a first-order quantum phase transition at the critical point $s=0$.
Here we showed that, in correspondence to the phase transition point, the  ground state undergoes a localization transition from completely delocalized to super-exponentially localized ($s>0$).
We showed how this ground state transition leads to a sharp change throughout the spectrum from a {\em fast} dynamical phase at $s<0$, where ergodicity is established quickly under unitary evolution, to a {\em slow} dynamical phase at $s>0$ where thermalization is impeded.
We provided rigorous results about the dynamical consequences of this transition focusing on the behavior in the slow non-thermalizing side. 

\paragraph*{Summary of the results.} In the following paragraphs we explicitly compile the main results of our work, as well as their connections to other models and possible generalizations. By combining analytical arguments, exact diagonalization and tensor network methods, we made the following findings.

(i) For a broad class of constrained models, we proved that the ground state is exponentially localized. In this class, the quantum East model is the simplest example.
The localized nature of the finite-size ground state for $s>0$ allows for a systematic construction of the ground state for arbitrary system sizes.
This construct is very simple, that of a tensor product of the ground state of a small system with a completely empty state.
Since the second factor is annihilated due to the constraints, all cost is concentrated at the juncture, which the localization in the first factor makes vanishingly small in the large size limit. 

(ii) This procedure can be extended to obtain exact large-size eigenstates of finite energy density.
By replacing the right factor by an excited state, one can systematically construct an exponential number of non-thermal excited states.
If the right factor is that of the eigenstate of maximal energy, the ensuing large-size eigenstate has area law entanglement.
This means that there are (at least) polynomially (in system size) many area law eigenstates of finite energy density.

(iii) By generalizing the tensor product construction to many junctions we can define an even larger class of non-thermal states in terms of what we call super-spins.
A state composed of super-spins is the tensor product of several ground states for a finite system of a fixed size, possibly separated by empty blocks of the same size, and thus corresponds to a dressed occupied spin localized at each occupied juncture.
From the arguments above, if the number of super-spins scales sub-extensively with system size, and the distance between junctions is large enough, such states become area-law eigenstates in the large size limit.
States with extensive number of super-spins in contrast, while may still have small energy variance, are not guaranteed to be eigenstates. 
These states are still provably slow since the evolution of all the correlations, observables, and entanglement entropy starting from them can be bounded by the magnitude of their energy variance.

Even if a generic state {\it may} still thermalize (i.e. we cannot claim non-ergodicity of the system in the thermodynamic limit), we proved that there exist an exponentially large family of product states ---
experimentally easy to prepare --- which retain long memory of initial conditions. They take exponentially long times to entangle a small sub-region and, in some cases, they do not thermalize at all. These are the states that underpin the slow dynamics of the model.

(iv) We performed extensive numerical results for small systems obtained with exact diagonalization, as well as for large systems using tensor networks.
In particular, we considered several quantities of interest, such as the entanglement entropy of the eigenstates, and the distributions of their last site occupation and of their maximal overlap with a product state.
The statistical analysis shows that many eigenstates exhibit atypical behavior, signaling the presence of non-thermal dynamical properties that go far beyond our analytical constructions.
These properties confirm for small sizes the singular change throughout the whole spectrum as one varies the parameter $s$ from negative to positive.
Although all our analytical constructions rely solely on the localization of the ground state, our numerical studies indicate that many other eigenstates have similar localization properties.
This suggests that the classes of non-thermal eigenstates and super-spin states that we discussed above, may be further extended by making use of localized excited states from small system sizes.

\paragraph*{Generality of the mechanism.} With the term localization we mean that the density of occupied sites is localized in the neighborhood of a certain position. What we uncover here is not limited to the quantum East model. Our findings reveal a general mechanism for a broad class of models that induces exponential localization to their ground state and consequently non-thermal dynamical features in the thermodynamic limit. In particular, in appendix~\ref{appdx:generalization_localization}, we proved exponential localization of ground states belonging to a class of models which includes textbook examples such as simple spin Hamiltonians with nearest neighbor interactions of the form $H=\sum_{i} J_{zz} \sigma^z_{i}  \sigma^z_{i+1}+J_{zx} \sigma^z_{i}  \sigma^x_{i+1}+J_{z} \sigma^z_{i}+J_{x}  \sigma^x_{i}$.

\begin{figure}
	\includegraphics[width=0.4\textwidth]{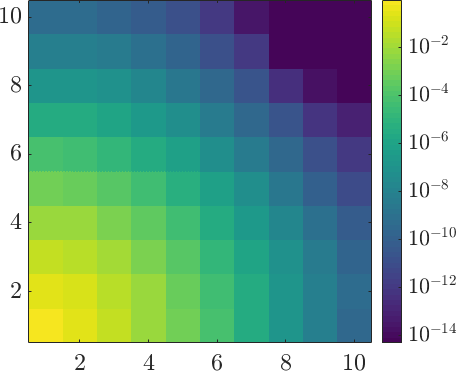}
	\caption{{\bf Single-site occupation of the quantum North-or-East model ground state with $s=0.01$ on a $10 \times 10$ lattice.}
	The low-entanglement structure allow to perform accurate DMRG calculations~\cite{Footnote1}.
  Even for small values of $s$, the ground state exhibits strong localization.
	}
	\label{fig:2Dgroundstate}
\end{figure}

\paragraph*{Extension to higher dimensions.} From classical KCMs \cite{Garrahan2018} we know that qualitative features are not very dependent on dimensionality.
It is natural to study the quantum generalizations of KCMs in higher dimensions, e.g. the quantum North-or-East model (see e.g. Ref.~\cite{Ashton_2006}), which extends the East model constraint to two dimensions.
Our DMRG calculation for the quantum North-or-East model~\cite{footnote2} on a $10\times 10$ lattice shows that the ground state remains localized even for small positive values of $s$ (see Fig.~\ref{fig:2Dgroundstate}).
This suggests that these two-dimensional KCMs should also display slow dynamics and a prominence of non-thermal states like the ones uncovered here for the quantum East model, therefore being candidates for disorder-free quantum systems exhibiting non-thermalization in higher dimensions.

\paragraph*{Comparison with other constrained models.} 

It is natural to compare the particular dynamical features we have discovered to those of other constrained models studied in the literature. 
First of all, the PXP model~\cite{Fendley2004,Lesanovsky2011}, a constrained system known to thermalize \cite{Ates2012} for typical states, exhibits a number of quantum scars --- excited states which fulfill the area law of entanglement. 
The PXP constraint allows spin-flips only when two nearest neighbors are in the down state, in contrast to the ``1-spin facilitation'' (albeit directional) of the quantum East model.
The PXP constraint is thus stronger than that of the East model, and as a consequence, in one dimension the state space breaks into exponentially many dynamically disconnected subspaces.
On the other hand, as we explained above, in the East model (with appropriate boundary conditions) all states are connected. 
Thus a {\em weaker} constraint in the quantum East model gives rise to {\em stronger} dynamical features, associated with the fact that regions devoid of occupied spins are locally frozen --- yet still dynamically connected --- a feature we exploited in the construction of non-thermal states. 
Notice also that the localization of the ground state that we have identified as crucial ingredient for the non-thermal features in the quantum East model, is not present in the scar states that inhibit thermalization~\cite{Turner2018} in the PXP model. As we showed above, it is in fact such localization that induces the presence an exponential number of scarred states in the system size and even the broader family of super-spin states.

A second class of models worth mentioning are those with fracton excitations, e.g.\  \cite{Chamon2005,Haah2011,Castelnovo2012,Yoshida2013,Prem2017,Nandkishore2018, Khemani2019Local,khemani2019localization,Rakovszky2020,Sala2020,Pretko2020}. In fact, this field of research started with Ref.~\cite{Chamon2005}, which generalised to the quantum realm pre-existing plaquette spin models of glasses \cite{Newman1999,Garrahan2000,Garrahan2002}, just like the quantum KCMs we study here are quantum generalisations of classical glassy KCMs.
As in the classical case, while there are connections between fracton models and quantum KCMs, there are some important differences. The models we consider have {\em explicit} kinetic constraints, while for fractons  
\cite{Chamon2005,Haah2011,Castelnovo2012,Yoshida2013} constraints are {\em effective}. 
This has significant consequences. Effective constraints are ``soft'' in the sense that they only partially prevent certain transitions, in contrast with the explicit ``hard'' constraints of the East model and its generalisations that cannot be broken. For example, this changes the nature of the phase transition of their ground states (which for some fracton models \cite{Devakul2019} can be inferred from the study of large deviations in the classical stochastic setting \cite{Turner2015}). Nevertheless, fracton models such as the Haah code (i.e.\ so-called type II \cite{Nandkishore2018,Pretko2020}) may display similar physics to that of the quantum East model. For example the heuristic arguments used in \cite{Castelnovo2012,Prem2017} to suggest super-Arrhenius relaxation at finite temperature in those models may also apply to the quantum East, since they posit a mechanism for relaxation which is the same as in the classical models. What would be more interesting is to explore whether the exact (in the thermodynamic limit) and fully quantum construction of excited states we present here can be translated in some manner to those models. 

More generally, it is important to distinguish the mechanisms for the emergence of non-thermal excited states --- and concomitant slow dynamics and potential non-ergodicity --- we have uncovered here from those based on what recently has been dubbed ``shattering of Hilbert space'' \cite{Prem2017,Nandkishore2018, Khemani2019Local, khemani2019localization, Rakovszky2020, Sala2020}. For the quantum East model and the boundary conditions we considered, see Sec.~\ref{sec:model}, the dynamics connects all the states in the Hilbert space. Furthermore, the non-thermal excited states that we find become exact {\em only in the limit $N \to \infty$}. That is to say that the change throughout the spectrum from $s<0$ to $s>0$ has the character of a true phase transition, not occurring at finite $N$ but in the thermodynamic limit.

Classically, the issue of fragmentation of configuration space for Markov generators of stochastic dynamics with constraints is well understood \cite{Ritort2003}. Before establishing whether a KCM dynamics is not ergodic, it is necessary to understand if the constraints make the generator {\em reducible}. Namely, whether there are regions of configuration space that are disconnected by the dynamics {\em at any finite system size}. When a dynamical generator is reducible, there can be an apparent breakdown of ergodicity simply by starting with an initial condition which has weight on disconnected sectors. However, reducibility should not be confused with non-ergodicity, which deals with diverging relaxation times within {\em a single connected irreducible sector}, 
and which occurs in the large size limit only. 
For the quantum case, similar considerations may apply. Our results for the quantum East model show the emergence of a non-thermal behavior, which becomes exact in the large size limit, within an irreducible sector.(For further discussion on these issues, see also \cite{Mondaini2018,Shiraishi_Mori_Reply2018}).

\section{Future directions}\label{sec:future_directions}

The slow dynamics of the East model is a {\em first-order} phenomenon --- cf.\ the transition in the ground state. It is a consequence of having spatial coexistence of two very different kinds of dynamics.
That is, since a region with no occupied spins is locally stable, it can only be relaxed starting from the interface with an active region.
While here we studied explicitly the spectral properties of the East model with open boundaries, we expect to find similar slow characteristics in the case of periodic boundaries, and for other one-dimensional models with similar constraints, such as the quantum ``2-spin facilitated'' Fredrickson-Andersen model (FA) \cite{Ritort2003,Hickey2016} (a 1-spin facilitated model but with a symmetric constraint).

Finding new and broader classes of non-thermal states might give insightful information about the emergent slow dynamics we observe for small system sizes and, more importantly, could address the question of ergodicity breaking.
Along the lines of our constructions, one possible direction includes the characterization of excited states for smaller sizes in order to promote them to fundamental building blocks for eigenstates at larger sizes in other systems. Their characterization may as well contribute to the understanding of the dynamical properties of their classical counterparts. 
The generalizations discussed above open a number of possible directions of investigation which include: (i) breakdown of ergodicity due to kinetic constrains for disorder-free models in one and higher dimensions; (ii) exploration of the corresponding dynamical transition and the concomitant singularity in the eigenstates; (iii) find physical implementations where these phenomena can be observed; (iv) extension of the results in appendix~\ref{appdx:generalization_localization} about the localization of ground states to higher dimensions and to models with different types of constraints.

\begin{acknowledgments}
The authors thank Johannes Feldmeier and Michael Knap for helpful and insightful discussions.
NP and GG especially thank Giuliano Giudici and Claudio Benzoni.
This work was partly supported by the 
Deutsche Forschungsgemeinschaft (DFG, German Research Foundation) under Germany's Excellence Strategy - EXC2111 - 390814868, and by the European Union through the ERC grants QUENOCOBA, ERC-2016-ADG (Grant no. 742102) 
, and through EPSRC Grant no.\ EP/R04421X/1.
NP acknowledges financial support from ExQM.
\end{acknowledgments}

\bibliography{library}

\begin{thebibliography}{105}
\bibitem [{foo()}]{footnote2}%
  \BibitemOpen
  \bibinfo {note} {The Hamiltonian of the quantum North-or-East model is
  defined as $$H_{2D} = -\frac{1}{2} \sum_{i,j} n_{i,j} \left[ e^{-s}
  \left(\sigma^x_{i+1,j} + \sigma^x_{i,j+1}\right) - \mathbb{1} \right].$$ The
  boundary conditions are chosen such that the $(1,1)$-site is pinned at $1$,
  and an additional field $n$ is placed on the opposite corner.}\BibitemShut
  {Stop}%
\end{thebibliography}

\appendix

\section{Localization of the ground state for generalized East models}\label{appdx:generalization_localization}

\begin{figure*}
\includegraphics[width=0.48\textwidth]{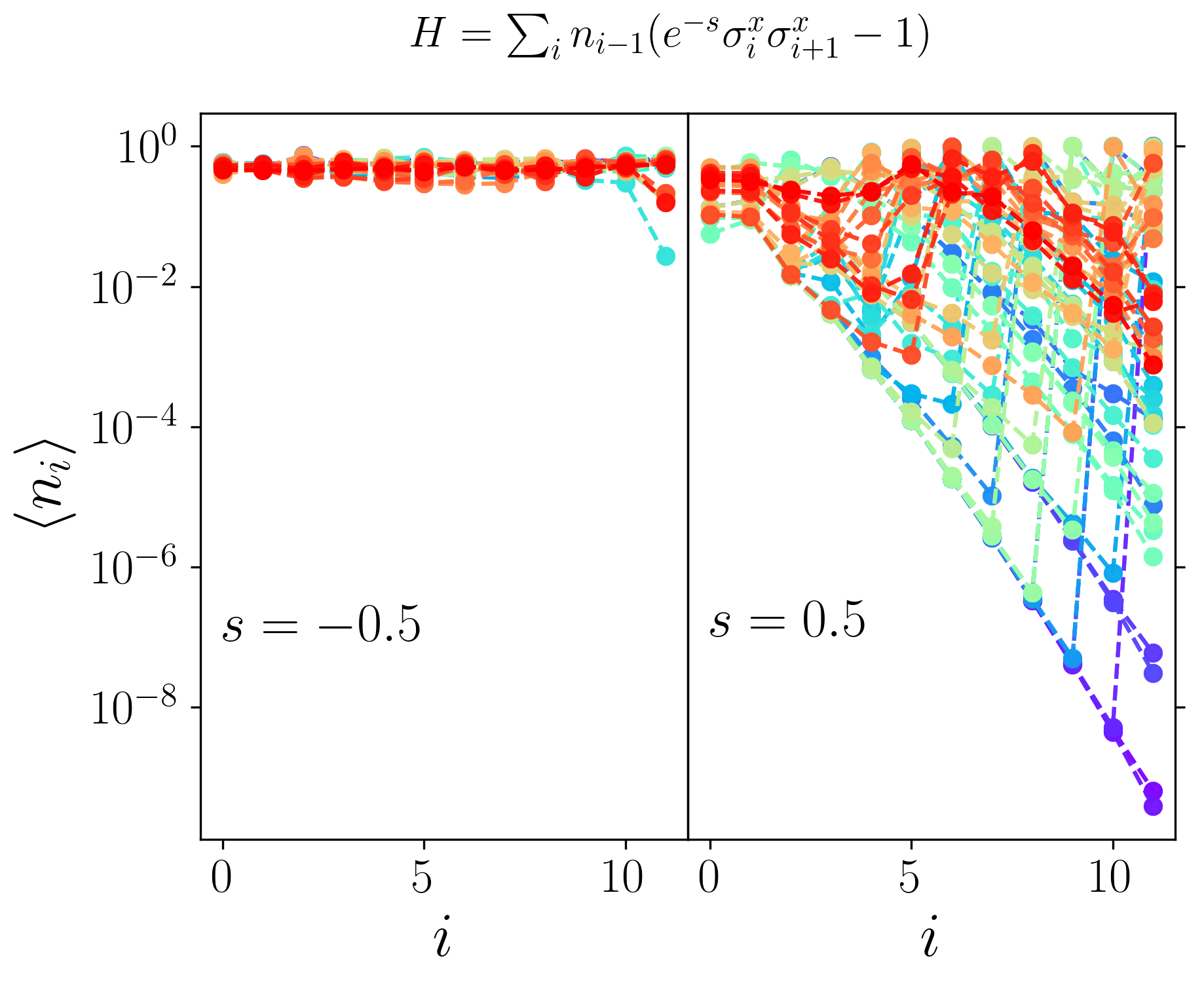}
\includegraphics[width=0.48\textwidth]{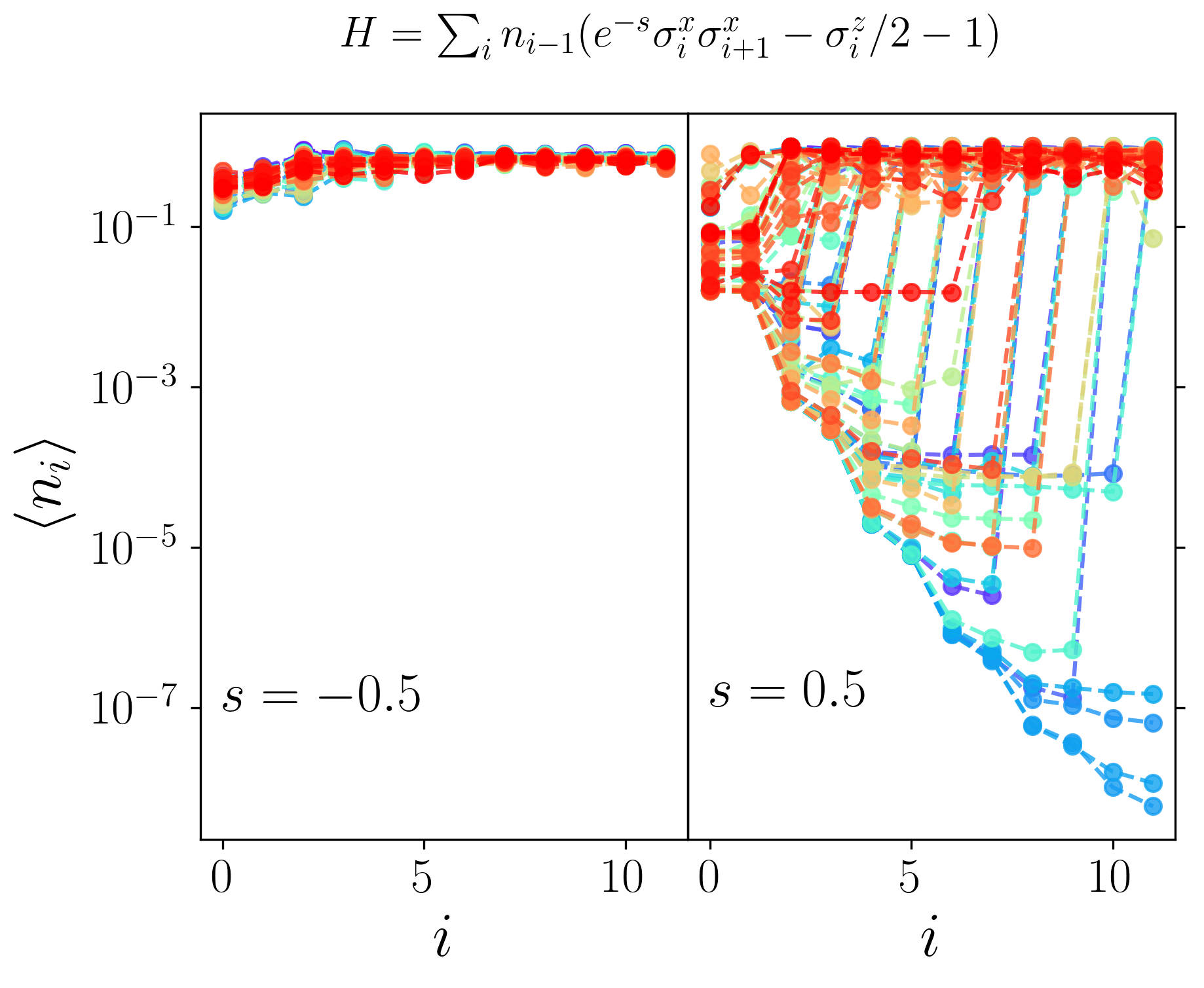}
\caption{{\bf Examples of other localized models.} As in Fig.~\ref{fig:ent_support} (right), we plot the expectation value of the occupation as a function of the support for two constrained Ising models. Both exhibit an exponential localization for positive values of $s$ and no localization for negative values ($N=12$). In the localized phase there are several other localized excited states. 
  }
  \label{fig:east_ising}
\end{figure*}

In this appendix we present 1D models that possess localized eigenstates. We consider a chain of qubits, with Hamiltonian
 \begin{equation}\label{eq:east_generalization}
 \tilde H= \sum_{i=-N}^N n_i (\tilde Q_{i} + \epsilon \tilde X_{i}) 
 \end{equation}

where $n_i=|1\rangle\langle 1|$, acting at position $i$, and $\tilde Q_{i}$ and $\tilde X_{i}$ act at positions $i+1,\ldots,i+L_{q,x}$, respectively, and fulfill the following conditions: (i) $\tilde Q_i$ is classical, in the sense that it commutes with all $n_j$, and has $|0\rangle^{\otimes L_x}$ as ground state (possibly degenerate) with energy $\tilde E_i$; (ii) $\tilde X_i$ is hermitian, bounded (i.e. $|| \tilde X_i||_{\rm op}\le c$ for some $c$), and its minimal eigenvalue, $x_i>-\tilde E_i$. We will show that there exists some $\epsilon_0>0$ so that if $\epsilon<\epsilon_0$, then $\tilde H$ has localized eigenstates in the thermodynamic limit $N\to\infty$. The set of Hamiltonians fulfilling those conditions include translationally invariant ones, where we can just take $\tilde Q_i$ and $\tilde X_i$ to be independent of $i$. Also, we have restricted this discussion to qubits, but the extension to higher dimensional systems is straightforward.

Let us first, without loosing generality, simplify the notation. We can take $L={\rm max}(L_x,L_q)$ and consider that both $\tilde Q_i$ and $\tilde X_i$ act on $L$ sites. We can also define $X_i=\tilde X_i-x_i \Id$ and $Q_i=\tilde Q_i + \epsilon x_i$. With these definitions, $X_i\ge 0$ and the lowest eigenvalue of $Q_i$ is $E_i=\tilde E_i + \epsilon x_i \ge E_{\rm inf} >0$ where
\begin{equation}
 E_{\rm inf}={\rm inf} E_i>0.
\end{equation}
Also, we can consider $||X_i||_{\rm op} \le 1$ just by defining $\epsilon\to \epsilon/c$. We will consider from now on $\epsilon < E_{\rm inf}$.

Notice that the class of Hamiltonians in Eq.~\eqref{eq:east_generalization} includes simple nearest neighbor spin systems by defining $\tilde Q_i = a \Id + b \sigma_i^z$ and $\tilde X_i = c \Id + d \sigma_i^x$. The quantum East model as described in the paper is 
one particular instance of this, with $a=(1- e^{-s})/2$, $b=0$, $c=1/2$, and $d=-1/2$.

The ground state of $\tilde H$ is the one with all qubits in $|0\rangle$ and has zero energy. Note that $H\ge 0$ and that whenever the $i$-th qubit is in state $|1\rangle$, it will add an energy at least $E_{\rm inf}>0$. 
As in the case of the quantum East model analyzed in the main text, a leading substring of $\ket{0}^{\otimes N}\ket{1}_0$ is preserved by the Hamiltonian, so that 
we can focus on eigenstates of the form

\begin{equation}
 |0\rangle^{\otimes N} \otimes |1\rangle_0 \otimes |\psi\rangle
\end{equation}
where $\ket{\psi}\in {\cal H}$ is a vector in the Hilbert space corresponding to the qubits $1,\ldots,N$. In the following, we will restrict the discussion to this Hilbert space, and we will be ultimately interested in the limit $N\to \infty$.
We now define subspaces $\mathcal{S}_0=|0\rangle^{\otimes N}$ and $\mathcal{S} _n= (\mathbb{C}^2)^{n-1}\otimes |1\rangle_n \otimes |0\rangle^{\otimes N-n}$, i.e. those that have a 1 at the $n$-th position and zeros to its right. Obviously, $\mathcal{H}=\oplus_{n=0}^N \mathcal{S}_n$. We also define $P_n$ as the 
projectors onto $\mathcal{S}_n$.

We say that a state $\psi\in \mathcal{H}$ is exponentially localized if there exists some $\lambda>0$ such that
\begin{equation}
|| P_n|\psi\rangle ||^2 < e^{-\lambda n}
\end{equation}
Note that this automatically implies that
\begin{equation}
 \sum_{n=n_0}^\infty || P_n|\psi\rangle ||^2 < k e^{-\lambda n_0}
\end{equation}
where $k=(1-e^{-\lambda})^{-1}$.

Now, we take the part of $\tilde H$ acting on the restricted space, and define 
\begin{equation}
 H = H_0 + \epsilon V,
\end{equation}
where
\begin{eqnarray}
 H_0 &=& (Q_0 -E_0) + \sum_{i=1}^N n_i Q_i, \\
 V &=& X_0 + \sum_{i=1}^N n_i X_i.
\end{eqnarray}
Here, we have separated the term at $i=0$ taking into account that the state of the qubit at that position is $|1\rangle_0$.
We have also subtracted the energy $E_0$. With that, the ground state of $H_0$ is $|0\rangle^{\otimes N}$ and has energy $E=0$. In addition, $H_0$ has a gap $\Delta=E_{\rm inf}>0$, which corresponds to a state with just one qubit in $|1\rangle$. Furthermore, according to the discussion above
\begin{equation}\label{Xi}
 0\le X_i \le \Id.
\end{equation}

We consider the ground state of $H$, fulfilling the eigenvalue equation
\begin{equation}\label{Schroe}
 H|\varphi\rangle = E |\varphi\rangle
\end{equation}
We will show now that
\begin{equation}
\Delta_0 = \Delta -  E >0
\end{equation}
and that $\ket{\varphi}$ is exponentially localized. The first statement is rather obvious, since $H\ge 0$ and we can just use the variational principle
\begin{equation}
 E \le \langle \Psi_0|H|\Psi_0 \rangle =\epsilon \langle \Psi_0|X_0|\Psi_0 \rangle \le \epsilon < E_{\rm inf}=\Delta,
\end{equation}
where $|\Psi_0\rangle = |0\rangle^{\otimes N}$ and we have used that $\langle \Psi_0|H_0|\Psi_0\rangle= \langle \Psi_0|X_i|\Psi_0\rangle=0$ for $i>0$.

In order to show that $\ket{\varphi}$ is localized let us take the limit $N\to \infty$ and project (\ref{Schroe}) onto $\mathcal{S}_n$ with $n\ge L$,
\begin{equation}
 H_n P_n |\varphi\rangle = - \epsilon \sum_{n\ne m=0}^{\infty} P_n V P_m|\varphi\rangle
\end{equation}
where $H_n=P_n H P_n -E$. Note that the RHS vanishes unless $n-L\le m \le n+L$, since otherwise $V$ does not connect the subspace $\mathcal{S}_m$ with $\mathcal{S}_n$. Furthermore, as $X_i\ge 0$,
\begin{equation}
\label{Hnbound}
 H_n\ge P_nH_0P_n\ge  \Delta_0 P_n
\end{equation}
since $H_0$ has a gap, $\Delta_0$, and $n>0$, so that $P_n$ projects onto a subspace that is orthogonal to the ground state $\mathcal{S}_0$. 
Thus, we can invert $H_n$ on $\mathcal{S}_n$,
\begin{equation}
P_n |\varphi\rangle = - \epsilon \frac{1}{H_n} \sum_{n\ne m=n-L}^{n+L} P_n V P_m|\varphi\rangle
\end{equation}
Taking the norm of both sides, denoting by $r_n=||P_n|\varphi\rangle||$, and taking into account (\ref{Hnbound}) and (\ref{Xi}), we obtain
\begin{equation}
 r_n \le \epsilon_1 \sum_{n\ne m=n-L}^{n+L} r_m,
\end{equation}
with $\epsilon_1=\epsilon L/\Delta_0$. Defining now
\begin{equation}
 R_k = \sum_{n=0}^{L-1} r_{kL+n}
\end{equation}
for $k=0,1,\ldots$, it is straightforward to show that for $k>0$
\begin{equation}
 \label{Rk}
 R_k \le \epsilon_2 (R_{k-1} + R_{k+1})
\end{equation}
where $\epsilon_2=\epsilon_1 L /(1-\epsilon_1 L)$.

Now, we will show that $R_k\le \mu R_{k-1}$ with $\mu=2 \epsilon_2/(1+\sqrt{1-4\epsilon_2^2})<1$, so that $R_k\le \mu^k$ since $R_0\le 1$ as the state $\varphi$ is normalized, and thus
\begin{equation}
 r_n \le \mu^{n/L}.
\end{equation}
For that, we use recursively (\ref{Rk}) to obtain
\begin{equation}
 \label{aux}
 R_k \le \epsilon_2 R_{k-1} \sum_{m=0}^t \epsilon_2^{2m} C_m + \epsilon_2^{2t+1} \sum_{m=0}^{t} R_{k+2m+1} d_m
\end{equation}
for any value of $t>0$. Here, $C_m=\frac{1}{m+1} {2m \choose m}$ are the Catalan numbers, 
which count 
all possible paths in which a walk on $\ell$ that starts at $\ell=1$ and increases or decreases $\ell$ by one unit at every step can reach $\ell=0$ in $2m+1$ steps, without visiting $\ell=0$.
Analogously, $d_m$ is the number of paths that start at $\ell=1$ and end at $\ell=2m+1$ in exactly $2m+1$ steps without visiting $\ell=0$. We can bound $d_m$ by the binomial number, which would count the number of paths without the restriction of not visiting $\ell=0$. Thus $d_m\leq {2t \choose 2m}\leq {2t \choose t} \leq 2^{2t}$.
For $\epsilon_2<1/2$,
the second term in (\ref{aux}) vanishes in the limit $t\to\infty$, while the first one gives us the desired bound.

As a demonstration, in Fig.~\ref{fig:east_ising}, we show the decay of the occupation for two constrained Ising models. For positive values of $s$, both models exhibit exponential localization as predicted by our analytical treatment. Whereas for negative values, all the eigenstates have a homogeneous occupation. As in the case of the quantum East model in the localized phase, there are additional localized states above the ground state.

\section{Super-exponential decay in the quantum East model}\label{appdx:perturbation_theory}
\begin{figure}
	\includegraphics[width=0.9\columnwidth]{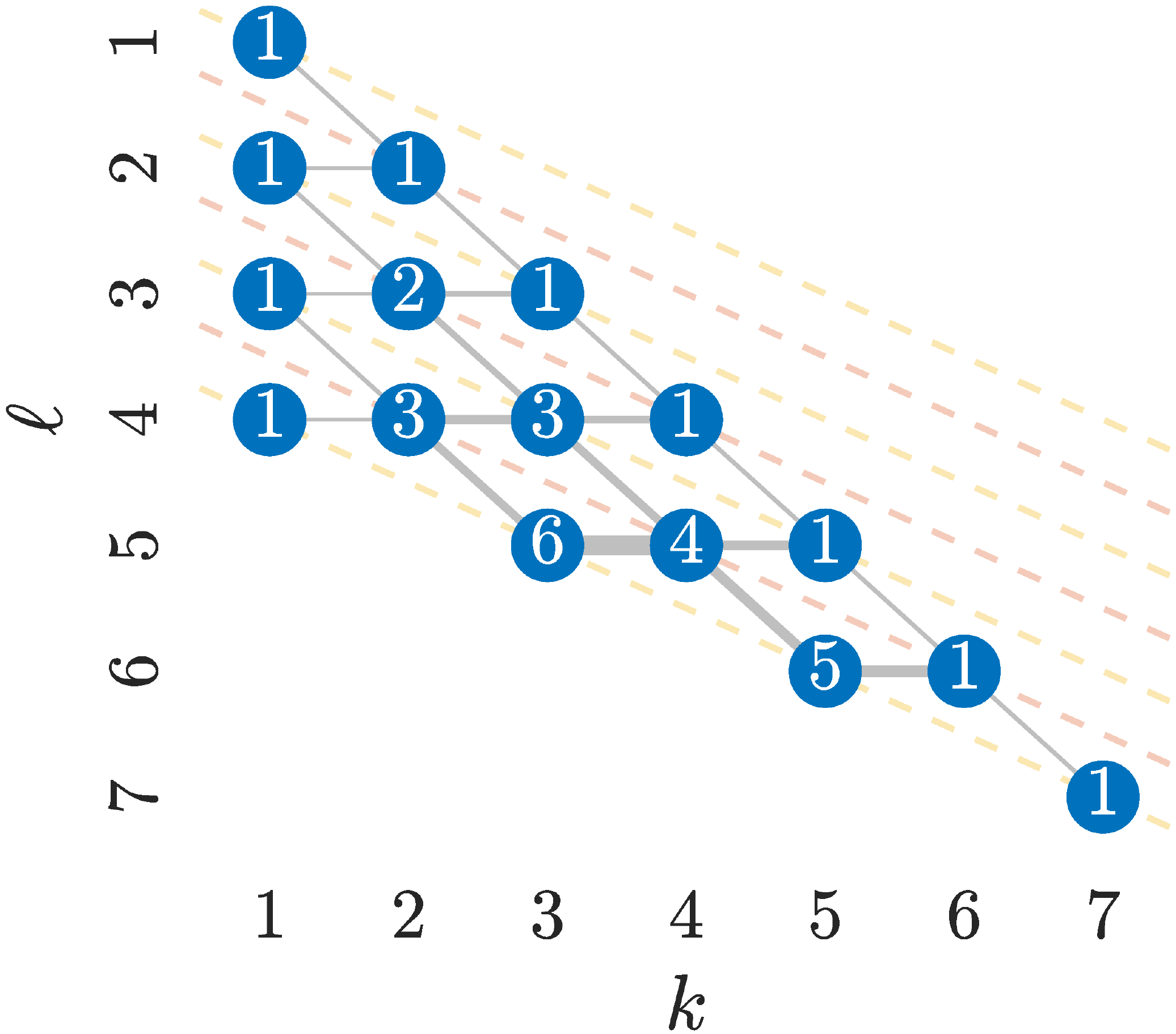}
	\caption{
	{\bf Transition graph of states in perturbation theory.}
	The states contained in the first seven orders of perturbation theory are illustrated as a graph, where each nodes groups states with the same $k$ (the number of occupied sites) and $\ell$ (the position of the rightmost occupied site).
	The number of states for fixed $(k,\ell)$ is indicated on each node.
	The thickness of each edge represents the number of transitions between states contained in the nodes via an application of $RV$.
	We notice that the graph has the structure of a Pascal triangle, where its shallow diagonals are highlighted with a dashed line.
	Notice that with an even (odd) number of hops along the edges  from the topmost node $\ket{ 1000\dots }$ we end up on an even (odd) shallow diagonal, colored in yellow (red).
	}
	\label{fig:perturbationgraph}
\end{figure}
We split Eq.~\eqref{eq:H+/-} into a bare Hamiltonian and an interaction term
\begin{equation}
  H^N_\pm  = \frac{1}{2} \left(H_0 + \varepsilon V_\pm \right) ,
\end{equation}
where
\begin{equation}
  \begin{split}
    H_0 &= \sum_{i=1}^{N-1} n_i \\
  V_\pm &= -  \sigma^x_1 -\sum_{i=1}^{N-1} n_i \sigma^x_{i+1} + (1 \mp e^{-s}) n_N .
  \end{split}
\end{equation}

We treat $\varepsilon = e^{-s}$ as a small parameter, so we can look at the first order corrections in $\varepsilon$ of the ground state deep in the localized phase, i.e. $s \to \infty$.
In the following, we consider only  the perturbation $V = V_+$, since the following calculations are the same for the other case.
The bare Hamiltonian $H_0$ is diagonal in the computational basis and has as unique ground state $\ket{ \chi^{(0)} } = \ket{ 000\dots }$ with a unit gap above it.
The excited states are labeled by $\ket{ k_\alpha }$, where $k=1,\dots,N$ counts the number of occupied sites and $\alpha$ labels the degeneracy: $H_0 \ket{ k_\alpha } = k \ket{ k_\alpha }$.

Using the notation in Ref.~\cite{Sakurai1994} we can express the corrections to the ground state as a Taylor expansion in the parameter $\varepsilon$
\begin{equation}\label{eq:perturbation_series}
  \begin{split}
    H^N_+ \ket{\chi} &= E_0 \ket{\chi}, \\
    \ket{\chi} &=  \ket{ \chi^{(0)} } +  \varepsilon \ket{ \chi^{(1)} } +  \varepsilon^2 \ket{ \chi^{(2)} } \dots , \\
    E_0 &= \frac{1}{2} \left( E^{(0)}  + \varepsilon E^{(1)} + \varepsilon^2 E^{(2)} + \dots\right) .
  \end{split}
\end{equation}
By defining the resolvent matrix
\begin{equation}
  R = - \sum_{k_\alpha} \frac{\ket{ k_\alpha } \bra{ k_\alpha }}{k} ,
\end{equation}
the $n$-th order terms can be computed as
\begin{equation} \label{eq:sakurai}
  \begin{split}
    \ket{ \chi^{(n)} } &= (R V)^n \ket{ \chi^{(0)} }, \\
    E^{(n)} &= \braket{ \chi^{(0)}  | V | \chi^{(n-1)} } .
  \end{split}
\end{equation}

From a more physical point of view, the perturbation should be seen as the introduction of a local impurity that has an effect that is localized at the boundary, rather than a global perturbation giving an extensive energy contribution.
This can be seen by looking explicitly the first orders in Eq.~\eqref{eq:sakurai}:
\begin{equation} \label{eq:perturbation_gs}
  \begin{split}
    \ket{\chi^{(1)}} &= \ket{ 1000\dots }, \\
    \ket{\chi^{(2)}} &= \frac{1}{2} \ket{ 1100\dots } ,\\
    \ket{\chi^{(3)}} &=\frac{1}{2} \left( \ket{  1000\dots } + \ket{ 0100\dots} + \frac{1}{3}\ket{ 1110\dots } \right) .
  \end{split}
\end{equation}
The corresponding energy is straightforward to compute
\begin{equation} \label{eq:perturbation_energy}
     E_0 = - \frac{\varepsilon^2}{2} - \frac{\varepsilon^4}{4} + \order\left(\varepsilon^6\right) .
\end{equation}

We now comment on the convergence  radius of the perturbation expansion.
Firstly, we notice that in a Banach space, if the series $\sum_{n=0} |\varepsilon^n| \| \chi^{(n)}\|$ converges, then the series for $\ket{\chi}$ converges too.
We shall now prove that the norm $\| \chi^{(n)} \|^2$ grows as the $n$-th power of the golden ratio $\varphi$.
In order to understand the structure of each perturbation theory order, it is instructive to construct a graph of transition between the states, as shown in Fig.~\ref{fig:perturbationgraph}.
The states are grouped in cluster labeled by the number $k$ of occupied sites and the length $\ell$ of the non-trivial string.
Each cluster contains $\binom{\ell}{k-1}$ states, and applying $RV$ to a state $\ket{ k,\ell}$ generates up to $k+1$ states in at most 4 distinct ways: $\ket{ k \pm 1,\ell }$ or $\ket{ k \pm 1,\ell \pm 1 }$.
Then the $n$-th perturbation term can contain at most the first $\lfloor \frac{n+1}{2} \rfloor$ even or odd ``shallow'' diagonals, depending on whether $n$ is even or odd.
In other words, the number of states generated by applying $RV$ to the $n$-th perturbation order is equal to the number of states in the order $(n-1)$ plus the number of states in the $n$-th shallow diagonal, which is the corresponding Fibonacci number $F_{n}$, since $\sum_{j=0}^{\lfloor \frac{n-1}{2} \rfloor}\binom{n-j-1}{j} = F_{n}$.
Hence the total number of states $\#^{(n)}$ in the $n$-th order is
\begin{equation}
  \#^{(n)} =
  \begin{cases}
    \sum_{j=0}^{\lfloor n/2 \rfloor} F_{2j+1} = F_{n+1} & n~\mbox{odd} \\
    \sum_{j=0}^{n/2} F_{2j} = F_{n+1} - 1 & n~\mbox{even} \\
  \end{cases}
\end{equation}
which asymptotically tends to $\varphi^{n+1}$.
The amplitude of each state will be the sum of all the paths of length $(n-1)$ leading to it from $\ket{1000\dots}$, weighted by $\prod_k 1/k$ of all the  visited states $\ket{ k_\alpha }$ along the path.
The weighted sum cannot exceed $(1/2)^{\lfloor (n-1)/2 \rfloor}$, since this is the path connecting $\ket{\chi^{(1)}}$ and $\ket{\chi^{(2)}}$.
Furthermore, the number of distinct paths cannot exceed $4^{n-1}$, since in principle  we can move in the 4 directions for $(n-1)$ moves, as shown in Fig~\ref{fig:perturbationgraph}.
While these estimates are not particularly tight, they are sufficient to bound the norm of $\ket{ \chi^{(n)} }$ as
\begin{equation}
  \| \chi^{(n)} \|^2 < \left(\frac{2^{2(n-1)}}{2^{\lfloor (n-1)/2 \rfloor}}\right)^2 \varphi^{n+1} \sim (2^3\varphi)^n
\end{equation}
Thus the perturbation series in Eq.~\eqref{eq:perturbation_series} is at least convergent in the radius
\begin{equation}
  \varepsilon < \frac{1}{2 \sqrt{2 \varphi}} \approx 0.278
\end{equation}
or equivalently $s \gtrsim 1.28$.
While this result is used to prove a positive radius of convergence, in practice we can monitor the rate of convergence of the norm of each order to have confidence in the method.
Indeed, numerically, we observe a rapidly converging norm in the regime $s \gtrsim 0.9$.

The perturbation theory picture not only gives us a method to compute the first order corrections to the ground state, but also an understanding of its structure.
Indeed, looking at the first terms in the expansion, we notice that the terms remain localized close to the left boundary.
As we apply powers of $V$, higher order terms in Eq.~\eqref{eq:sakurai} become progressively delocalized, but these contributions get damped at least by a factor $\varepsilon^n$.

\begin{figure*}
  \includegraphics[width=0.6\textwidth]{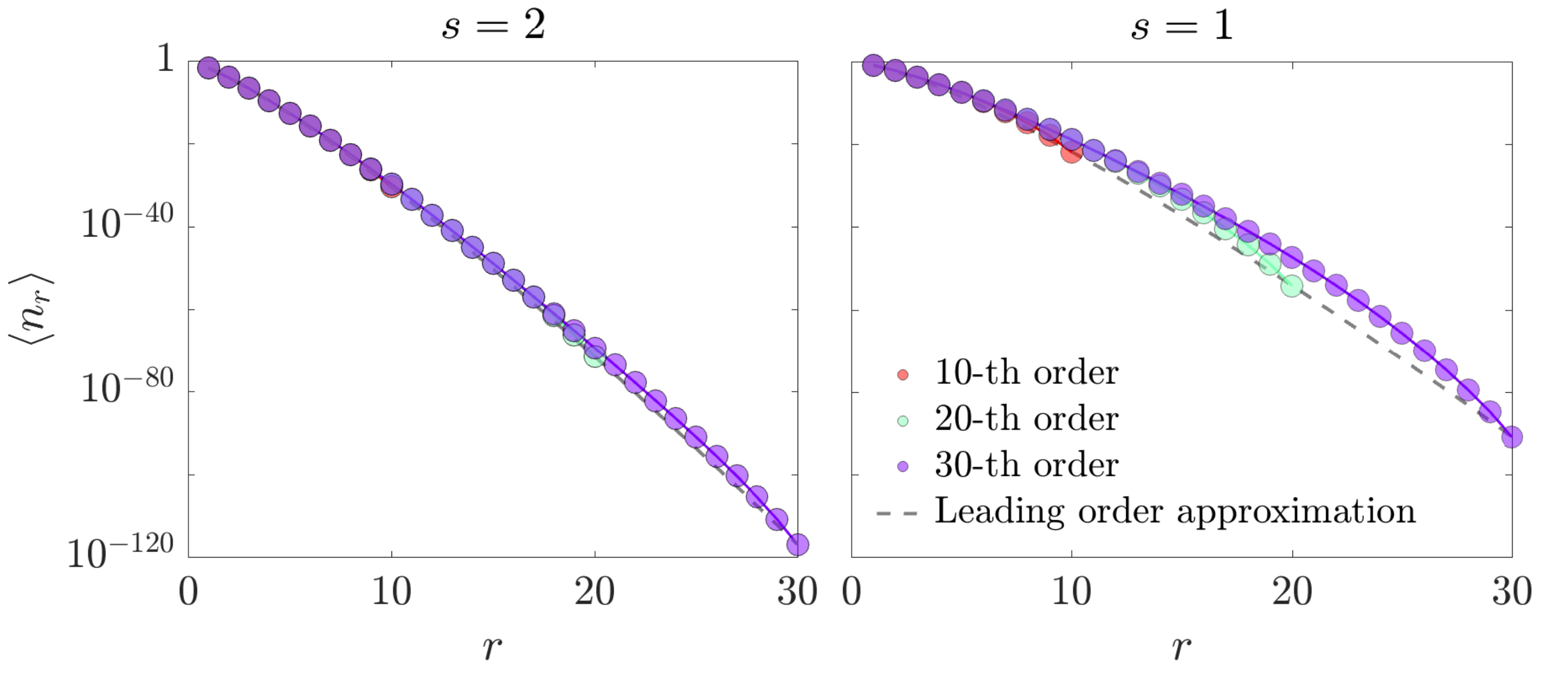}
	\caption{{\bf Perturbation theory results for the occupation profile.} The results for the occupation obtained using finite order perturbation theory and the extracted leading order contribution are illustrated for different values of the perturbation.
  We notice that, in the limit of weak perturbation, the two methods give similar results.
  Close to the boundary of the convergence radius, higher order term contributions become relevant, giving corrections only at larger distances.
  }
	\label{fig:perturbationdelta}
\end{figure*}

Let us define the domain-wall state $\ket{ \Theta_m } = \ket{ 1 }^{\otimes m} \ket{ 000 \dots }$.
From Eq.~\eqref{eq:perturbation_gs}, we notice that $\braket{ \Theta_n | \chi^{(n)}} = 1/n!$ and that $\braket{ \Theta_{m>n} | \chi^{(n)}} = 0$.
At each order $n$ there is a contribution $\ket{ \Theta_n }$ coming from a $\ket{ \Theta_{n-1} }$ in the previous order.
In Fig.~\ref{fig:perturbationgraph}, these states correspond to the rightmost diagonal, i.e. $k = n$.
One can think of this term as the fastest possible ``excitation'' created by applying $V$ repeatedly.
Since all other terms in $\ket{ \chi^{(m)} } $ have a smaller support, clearly $n_m \ket{ \chi^{(m)} }  = 1/m!$, and more generally, at the first order
\begin{equation}
  n_r \ket{\chi} = \frac{\varepsilon^r}{r!} \ket{ \Theta_r } + \dots .
\end{equation}
We can then conclude that
\begin{equation}\label{eq:perturbation_localization}
  \braket{ n_r } = \frac{\braket{ \chi | n_r | \chi }}{\braket{ \chi | \chi }} = \left( \frac{\varepsilon^r}{r!} \right)^2 + \order\left(\varepsilon^{2r + 1}\right) .
\end{equation}
We have dropped the normalization factor, since $\braket{\chi|\chi} = 1 + \order\left( \varepsilon^2 \right)$ leads to sub-leading corrections.
Checking with finite-order expansions, Eq~\ref{eq:perturbation_localization} gives a qualitatively accurate prediction of the decay of the occupation number, as shown in Fig.~\ref{fig:perturbationdelta}.
Using Stirling's formula, the asymptotic behavior of  Eq.~\eqref{eq:perturbation_localization} at large distances is $\braket{ n_r } \sim \exp[-2r \log{r} + 2r(1-s)]$.
Therefore, in the perturbative regime, the ground state exhibits super-exponential localization around the first site.

These perturbation calculations can equivalently be connected to the adiabatic theorem~\cite{FetterWalecka}.
By introducing a time-dependent coupling $\varepsilon(t)$, the different perturbation orders correspond to the expansion of the evolution operator.
Similar perturbation calculations were performed on the quantum East model in the case of periodic boundary conditions~\cite{Crowley2017}, in order to derive an effective Hamiltonian describing the hopping of domain wall in a given magnetization sector.

\section{Computation of the variance}\label{appdx:comp_var}
Consider the Hamiltonian defined in Eq.~\eqref{eq:QEast} on a one dimensional lattice of $N$ sites, and the eigenstates,
\begin{equation}
  \begin{split}
    \ket{ \Phi_{\pm} } = & \ket{ \phi _{\pm} (N-1) } \otimes \ket{ \pm }, \\
    H ^N \ket{ \Phi_{\pm} } = & H_{\pm}^{N-1} \otimes \Pi_{\pm}  \ket{ \phi _{\pm} (N-1) } \otimes \ket{ \pm } = E_{\pm} \ket{ \Phi _{\pm} }.
  \end{split}
\end{equation}
We consider the normalized state
\begin{equation}\label{eq:approx_eig_appendix}
  \begin{split}
    \ket{ \Psi_+ } = &  \ket{ \phi _{+} (N-1) } \otimes \ket{ 0 } ^{\otimes (L-N+1)} \\
    = & \sqrt{2} \Pi_0^N \ket{ \Phi_+ } \otimes \ket{ 0 } ^{\otimes (L-N)},
  \end{split}
\end{equation}
in the $\ket{+}$ sector, similar considerations hold in the $\ket{-}$ sector.
The state $\ket{\Psi_+}$ admits a simple computation of its energy expectation, and its variance.
In order to compute the variance, we need two ingredients.
The expectation value of the energy
\begin{equation}
  \braket{ \Psi_+ | H^L | \Psi_+ } = \Braket{ \Psi_+ | \left[ H_+^{N-1} \otimes \Pi_+ + H_-^{N-1} \otimes \Pi_- \right] |  \Psi_+ },
\end{equation}
which can be computed from
\begin{equation}
  \begin{split}
    \braket{ \Psi_+ | H_+^{N-1} \otimes \Pi_+  | \Psi_+ } = & \frac{1}{2} E_+ \\
    \braket{ \Psi_+ | H_-^{N-1} \otimes \Pi_-  | \Psi_+ }
    = & \frac{1}{2} \left( E_+ + e^{-s} \braket{ n_{N-1} }_{\phi_+} \right),
  \end{split}
\end{equation}
and takes the form
\begin{equation}
  \braket{ H^L } = \braket{ \Psi_+ | H^L | \Psi_+ } = E_+ + \frac{1}{2} e^{-s} \braket{n_{N-1} }_{\phi_+}.
\end{equation}
The expectation value of the square of the Hamiltonian is
\begin{equation}
  \begin{split}
    \braket{ H^L H^L } = & \Braket{ (H_+^{N-1})^2 \otimes \Pi_+ } \\
    + & \Braket{ (H_-^{N-1})^2 \otimes \Pi_- },
  \end{split}
\end{equation}
where we can explicitly calculate
\begin{equation}
  \begin{split}
    \braket{ (H_+^{N-1})^2 \otimes \Pi_+ } = & \frac{1}{2} E_+^2 \\
    \braket{ (H_-^{N-1})^2 \otimes \Pi_- } = & \frac{1}{2} \left(E_+^2 + e^{-2s} \braket{ n_{N-1} }_{\phi_+} \right) \\
    & + E_+ e^{-s} \braket{ n_{N-1} }_{\phi_+} .
  \end{split}
\end{equation}
We have now all the ingredients to compute the variance
\begin{equation}\label{eq:variance_appendix}
  \braket{ H^L H^L } - \braket{ H^L }^2 =  \frac{e^{-2s}}{2} \left[ \braket{ n_{N-1} }_{\phi_{\pm}} - \frac{1}{2} \braket{ n_{N-1} }^2_{\phi_{\pm}} \right] .
\end{equation}

\end{document}